\documentclass[aps,prb,reprint,superscriptaddress,amsmath,amssymb, groupedaddress]{revtex4-2}

\usepackage{graphicx}
\usepackage{dcolumn}
\usepackage{bm}
\usepackage{braket}
\usepackage{hyperref}
\hypersetup{
    colorlinks=true,
    linkcolor=blue,
    citecolor=red,
    filecolor=magenta,      
    urlcolor=blue,
}
\usepackage{cleveref}
\usepackage{amsthm}
\newtheorem{definition}{Definition}

\newtheorem{proposition}{Proposition}

\usepackage{xcolor}

\newcommand{\x}{\mathbf{x}}
\renewcommand{\k}{\mathbf{k}}
\newcommand{\y}{\mathbf{y}}
\renewcommand{\u}{\mathbf{u}}

\newcommand{\h}{\boldsymbol{h}}
\newcommand{\g}{\boldsymbol{g}}
\newcommand{\z}{\boldsymbol{z}}
\newcommand{\m}{\boldsymbol{m}}

\newcommand{\btheta}{\boldsymbol{\theta}}

\begin{document}
\title{Spectral Born machines: classically trainable quantum generative models for discrete data}
\author{Austin Huang}
\author{William Maxwell}
\author{Vasilis Belis}
\author{Evan Peters}
\author{Jason Pye}
\author{Soran Jahangiri}
\author{Joseph Bowles}
\email{joseph@xanadu.ai}
\affiliation{Xanadu Quantum Technologies Inc., Toronto, Ontario, M5G 2C8, Canada}
\date{\today} 

\begin{abstract}
We present \emph{spectral Born machines}, a class of quantum generative models that results from viewing and generalizing the class of IQP Born machines through the lens of group Fourier analysis. These quantum models exploit the quantum Fourier transform to create an inductive bias that make them naturally suited to learning integer-structured data, while remaining classically hard to sample from in general. Similar to IQP Born machines, spectral Born machines can be trained efficiently at scale on classical hardware via a maximum mean discrepancy loss based on graph spectral analysis, which we make available in a new \emph{tcdq} module of the PennyLane software platform. In numerical experiments, we show how the spectral bias of the model leads to significantly reduced parameter counts compared to unstructured approaches, and demonstrate the scalability of the software by training a 190-qubit model with over 1 million parameters to successfully learn a distribution of 93 nucleotide-long ribosomal RNA. Our results suggest that highly over-parameterized spectral Born machines may be immune to overfitting, even in strongly data-scarce regimes.
\end{abstract}

\maketitle


\section{Introduction}
The ability to use interference to construct and sample from high-dimensional probability distributions is what enables quantum computers to surpass the limits of classical computation. This is closely tied to the task of generative machine learning, where one aims to construct---or learn---a probability distribution given access to a finite set of empirical data. Viewed from this angle, generative machine learning is a natural area in which quantum computers may provide advantages for machine learning. 

Modern machine learning models invariably make use of a large number of trainable parameters \cite{kaplan2020scaling}. This is beneficial since it allows us to represent the belief that the ground truth distribution is a complicated, unknown function, while simultaneously adding a soft bias towards `simpler' distributions that generalise better due to the benefits of scale \cite{wilson2025deep, wilson2020bayesian,liu2022loss}. It is therefore  likely that successful quantum generative models will require at least modestly large parameter counts. However, even with an optimistic view on the clock speeds of future quantum computers \cite{hoefler2023disentangling}, it is now well understood that traditional approaches to quantum circuit gradient estimation are  unable to scale beyond relatively small numbers of parameters. New approaches to training are thus in desperate need \cite{bowles2025backpropagation, abbas2023quantum, coyle2025training, chinzei2025trade, mathur2026scalable, coyle2026adaptive}.

One solution to this problem was recently proposed in the form of a `train classical, deploy quantum' (TCDQ) approach \cite{recio2025train, kasture2023protocols}. Here, the basic idea is to construct a highly parameterized generative model based on a class of quantum circuits that are believed to be hard to classically sample from, while also identifying classically simulable observables that can be used to train the model. Since the observables are classically simulable, training can be performed on classical hardware. A quantum computer is therefore only required at inference time in order to sample from the model, where genuine computational advantages may be present. This enables scalable training, dramatically reducing the amount of quantum resources needed in the overall pipeline, while avoiding quantum data-loading bottlenecks since the model only sees data during the classical training loop. The possibility of training quantum generative models at scale in this way was first shown for instantaneous quantum polynomial (IQP) Born machines \cite{recio2025train}, and later for fermionic \cite{bako2025fermionic}, bosonic \cite{gottlieb2026efficient, kolarovszki2026generative, kurkin2026universality} and forrelation-based \cite{tuysuz2026quantum} circuits, and has attracted considerable attention as a promising route to quantum advantage \cite{herrero2025born,ballo2026shallow, herbst2025limits, slim2026iqp, baumann2026parity}.

\begin{figure*}
    \centering
\includegraphics[width=\textwidth]{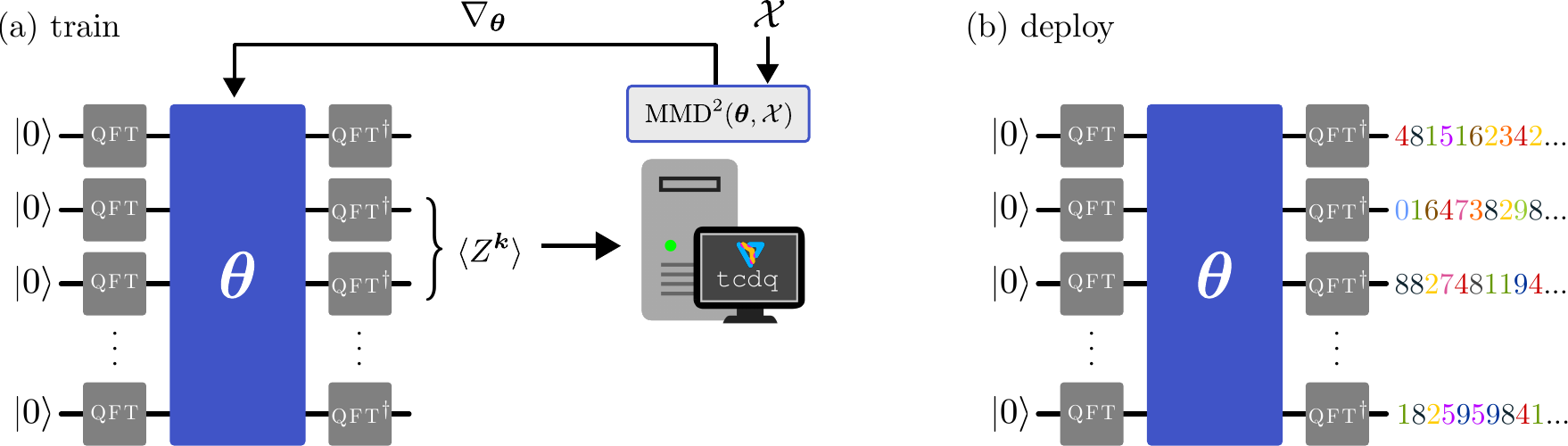}
    \caption{(a) A spectral Born machine consists of a diagonal parameterized unitary sandwiched between two layers of quantum Fourier transforms acting on qudits. For these circuits, expectation values of Heisenberg-Weyl observables can be classically estimated in order to train the maximum mean discrepancy loss function between the output distribution of the circuit and the empirical data distribution, which we make available via the PennyLane module \texttt{tcdq}. (b) Once training is complete, the trained circuit can be deployed on quantum hardware and used to sample new data, where genuine computational advantages may be present.}
    \label{fig:placeholder}
\end{figure*}
All the existing models are however natively binary, meaning they are designed to learn distributions over bitstrings. This presents a problem, since most datasets of interest are not binary in nature, and as we explain below, simply encoding higher dimensional data via a binary encoding is not ideal due to the mismatch between the natural structure of the integers and bitstrings. The main drive of this work is therefore to construct a TCDQ-type generative model that is designed specifically with the discrete structure of integers in mind. To do this, we look to group theory and Fourier analysis, which provide the mathematical tools to reason about both structure and smoothness for such data. In particular, since we are working with data consisting of vectors of discrete values, the group we consider is $\mathbb{Z}_d^n$, the $n$-fold product group of the group $\mathbb{Z}_d$ that describes the integers $\{0,\cdots,d-1\}$. 

As we will see, the original IQP Born machines can be understood as models that parameterize the phase structure of the quantum state in Fourier space. Since IQP Born machines are based on qubits, the relevant Fourier transform is the group Fourier transform over $\mathbb{Z}_2^n$, which is realized by a layer of Hadamard unitaries. If we replace the Hadamard unitaries in an IQP Born machine by quantum Fourier transforms over the group $\mathbb{Z}_d$, and define an analogous parameterized diagonal phase layer, we obtain a classically trainable model that ties directly to the group-Fourier structure of $\mathbb{Z}_d^n$. Importantly for machine learning, this structure allows us to control inductive biases of the generative model distribution that can be expected to aid generalization. On the one hand, we can directly bias towards distributional smoothness \cite{rosca2020case, belis2026spectral, dherin2022neural} with respect to the ordering of integers by selecting trainable gates that control only the low-frequency content of the phase unitary. On the other hand, we can bias the training of the model to match the structure of either ordinal or categorical discrete variables, which we show by formulating two versions of the maximum mean discrepancy loss function based on graph spectral analysis.

Like IQP Born machines, we show how training can be efficiently implemented using basic linear algebra operations, which allows the method to scale to large parameter and qudit counts. To facilitate future research, we implement the method in a dedicated \texttt{tcdq} module \cite{tcdq_note} of the PennyLane software platform \cite{bergholm2018pennylane}. The software is written in JAX \cite{jax2018github}, making algorithm construction flexible, differentiable, and scalable. Using \texttt{tcdq}, we first train a spectral Born machine on synthetic data generated via the Potts model. This distribution has a bias towards smoothness over the integers, making it a natural testbed for our model class. We show that spectral Born machines with gate sets that are engineered to match this bias are able to successfully learn from the data using much fewer parameters than unstructured equivalents. Second, in order to demonstrate the software's scalability, we train a spectral Born machine with over 1 million parameters on a scarce distribution of 93-nucleotide ribosomal RNA, showing that it can accurately capture the low-body information. Although we cannot sample from the model, we argue that despite having a large number of parameters, the model does not appear to overfit to the data. This provides evidence that, despite extreme overparameterization, spectral Born machines may be immune to overfitting, even in strongly data-scarce regimes. 

Overall, we hope this work ignites further interest in classically trainable generative models, enables an exciting new era of empirical quantum machine learning, and ultimately leads to the discovery of groundbreaking new applications of quantum machine learning.\\[5pt] 

\paragraph*{Note.} During preparation of this manuscript we became aware of another work~\cite{banks2026qudit}, which generalizes IQP Born machines in a similar manner. We view the two works as complementary: Ref.~\cite{banks2026qudit} emphasizes an application to high-energy physics, with symmetry-adapted circuits, hardware compilation, and small-scale sampling-based validation, whereas this work develops the group-Fourier formulation over $\mathbb{Z}_d^n$, using general Heisenberg-Weyl observables and a graph-spectral MMD, discusses the associated spectral inductive bias, and releases software that we use to train models with over a million parameters.

\subsection{The problem of encoding integers to binary}\label{sec:binary}

Before describing our generative models, it is instructive to understand why we don't encode integer data to binary and learn the distribution in this space using existing models. For example, given data $\x\in\mathbb{Z}_d^n$, we could encode each of the $d$ integers directly into $\log_2(d)$ bits, using an encoding (for d=8) such as $\{0,1,\cdots,7\}\rightarrow\{000,001,\cdots,111\}$. However, being designed around the structure of bitstrings, existing quantum generative models are generally biased towards learning smooth functions with respect to the distance on the Hamming cube, and this implies a structural inductive bias that does not match that of the integers. More specifically, these models are typically built in such a way that the string $01111$ is designed to be far from the string $10000$, with a maximal (Hamming) distance 5. In the standard encoding from integers to bitstrings however, these two strings represent the numbers 31 and 32 and have minimal distance. The result of this mismatch between distances means that models that are smooth on the Hamming cube will not necessarily be smooth on the encoded integers and vice-versa, which given the importance of smoothness for learning \cite{rosca2020case, dherin2022why} is likely to hinder generalization. This problem can be mitigated to some extent via the use of Gray-codes \cite{krebsbach2026encoding}, but ultimately there is no escaping the fact that integers have exactly two neighbors while bitstrings have $n$.

Another approach, which is common in classical machine learning, is to encode integers via a one-hot encoding, for example mapping $\{0,1,2,3\}$ to $\{1000,0100,0010,0001\}$. The first obvious downside of this is that it requires $d$ qubits rather than $\log_2(d)$. More problematically however, in order to stay in the valid encoding subspace, the generative model must parameterize a distribution that outputs a bitstring that is a concatenation of Hamming weight 1 bitstrings, and, since one-hot bitstrings are equidistant, must encode additional structure to represent integer ordering. This effectively requires building integer structure around the choice of encoding, which is precisely what is achieved by spectral Born machines using fewer qubits.

\section{Fourier-based unitaries and observables}\label{sec:fp_circuits} 
We begin by describing the unitaries and observables that form the basis of our generative model. Both of these are natural generalizations of IQP unitaries and Pauli observables, obtained by working with the Fourier transform over $\mathbb{Z}_d^n$. 

\subsection{The Fourier transform over $\mathbb{Z}_d^n$}\label{sec:qft_zd}

The first ingredient we need is the group Fourier transform. For a finite abelian group $G$, the Fourier transform \cite{kondor2008group} of the complex-valued function $f:G\to\mathbb{C}$ and its inverse are given by 
\small
\begin{align}\label{eq:ft}
    \hat f(\k) :=& \sum_{\x\in G} f(\x)\,\chi_{\k}(\x), \quad f(\x) =& \frac{1}{|G|}\sum_{\k\in G}\hat f(\k)\,\chi^*_{\k}(\x),
\end{align}
\normalsize
where the group characters $\chi_\k:G\to\mathbb{C}$ correspond to the one-dimensional irreducible representations of $G$. 
We work with the product cyclic group $G = \mathbb{Z}_d^n$, which naturally relates to vectors of integers, for which the characters are given by $\chi_{\k}(\x) = \omega^{\k\cdot\x}$, with $\omega=\exp(2\pi i/d)$ the $d^{th}$ root of unity.


The corresponding quantum Fourier transform (QFT) \cite{childs2010quantum} can be realised via a tensor product of quantum Fourier transforms of $\mathbb{Z}_d$. Concretely, the QFT over $\mathbb{Z}_d$ and its inverse are
\begin{align}
    F\ket{x} &= \frac{1}{\sqrt{d}}\sum_{k} \omega^{kx}\ket{k}, \\
    F^{\dagger}\ket{x} &= \frac{1}{\sqrt{d}}\sum_{k} \omega^{-kx}\ket{k},
\end{align}
so that the quantum Fourier transform over $\mathbb{Z}_d^n$ and its inverse are
\begin{align}
    F^{\otimes n}\ket{\x} &= \frac{1}{\sqrt{d^n}}\sum_{\k\in\mathbb{Z}_d^n}\omega^{\k\cdot\x}\ket{\k}, \\
    (F^{\dagger})^{\otimes n}\ket{\x} &= \frac{1}{\sqrt{d^n}}\sum_{\k\in\mathbb{Z}_d^n}\omega^{-\k\cdot\x}\ket{\k}.
\end{align}
Note that for $d=2$, we have $F=H$ and so recover the transform $H^{\otimes n}$ found in IQP circuits. 

\subsection{Heisenberg-Weyl observables}\label{sec:hw_obs}
Heisenberg-Weyl (HW) observables \cite{hwobs} provide the natural unitary generalisation of the Pauli observables to the qudit setting and have close connections to the Fourier transform over $\mathbb{Z}_d$. Define the generalised `shift' and `clock' operators $X$ and $Z$ on a single qudit by their action on the computational basis,
\begin{align}\label{eq:clockphase}
    X\ket{x} = \ket{x+1\!\mod d}, \qquad Z\ket{x} = \omega^{x}\ket{x}.
\end{align}
The HW observable with parameters $(k,m)\in\mathbb{Z}_d \times \mathbb{Z}_d$ is
\begin{align}
    \mathcal{D}(k,m) = Z^k X^m e^{-i\pi km/d},
\end{align}
HW observables are unitary however not Hermitian. We can however define the Hermitian counterparts
\begin{align}\label{eq:Q_def}
    \mathcal{Q}(k,m) = \chi\,\mathcal{D}(k,m) + \chi^*\,\mathcal{D}^{\dagger}(k,m),
\end{align}
where $\chi = (1+i)/2$. Like the qubit Pauli observables, the $\mathcal{Q}(k,m)$ form a complete orthonormal basis of Hermitian operators with respect to the Hilbert–Schmidt inner product \cite{hwobs}. Conjugation by the Fourier transform sends 
\begin{align}\label{eq:F_conjugation}
    F\,\mathcal{D}(k,m)\,F^{\dagger} &= \mathcal{D}(m,-k), 
\end{align}
which is the qudit analogue of the qubit relation $H Z H = X$. 
On $n$ qudits we work with the tensor-product operators
\begin{align}\nonumber
    \mathcal{D}(\k,\m) = \bigotimes_{i=1}^n \mathcal{D}(k_i, m_i), \quad \mathcal{Q}(\k,\m) = \bigotimes_{i=1}^n \mathcal{Q}(k_i, m_i),
\end{align}
indexed by pairs $(\k,\m)\in\mathbb{Z}_d^n\times\mathbb{Z}_d^n$. Note that from \eqref{eq:ft} for a state $\ket{\psi}$ with computational basis distribution $p(\x)=|\langle \x \vert \psi \rangle|^2$, we have
\begin{align}\label{eq:autoconv}
    \bra{\psi} \mathcal{D}(\k,\boldsymbol{0}) \ket{\psi} = \hat{p}(\k).
\end{align}
The $\mathcal{D}(\k,\boldsymbol{0})$ observables thus describe the measurement that returns the Fourier coefficients of the computational basis measurement distribution of $\ket{\psi}$.

\subsection{Class of parameterised unitaries}\label{sec:circuit_class}
We consider unitaries of the form
\begin{align}
    U(\btheta) = (F^{\otimes n})^{\dagger}\,D(\btheta)\,F^{\otimes n},
\end{align} 
where $D(\btheta)$ is a diagonal parameterized phase unitary on $n$ qudits. Concretely, we require $D(\btheta)$ to act on the computational basis as
\begin{align}\label{eq:Dunitary}
    D(\btheta)\ket{\z} = e^{i\Phi_{\btheta}(\z)}\ket{\z}, \qquad \z\in\mathbb{Z}_d^n,
\end{align}
for some real-valued phase function $\Phi_{\btheta}:\mathbb{Z}_d^n\to\mathbb{R}$ that depends on a set of trainable parameters $\btheta=(\theta_1,\cdots,\theta_m)$. We call such unitaries \emph{Fourier-phase unitaries} and define them as follows.
\begin{definition}[Fourier-phase unitary]\label{def:fpcircuit}
    A \emph{Fourier-phase unitary} on $n$ qudits of dimension $d$ is a unitary transformation comprised of the following:
    \begin{enumerate}
        \renewcommand{\labelenumi}{(\roman{enumi})}
        \item A Fourier transform layer $F^{\otimes n}$ over $\mathbb{Z}_d^n$.
        \item A parameterised diagonal phase layer $D(\btheta)$.
        \item An inverse Fourier transform layer $F^{\otimes n\,\dagger}$.
    \end{enumerate}
\end{definition}
We further define a \emph{tractable} Fourier-phase unitary as a Fourier-phase unitary in which the phase function $\Phi_{\btheta}(\z)$ is efficiently computable by a classical algorithm for any choice of $\btheta, \z$, that is, in $\operatorname{poly}(n,d)$ time and space. We will see explicit examples of such functions in Sec.~\ref{sec:hw_phase}. 
 

\section{Classical trainability}\label{sec:classical_train}
Here we will show that for an initial state $\ket{0}$ evolved under a tractable Fourier-phase unitary, expectation values of Heisenberg-Weyl observables can be estimated to inverse polynomial additive precision by an efficient classical algorithm. This result generalises the classical estimability of Pauli-$Z$ expectations for parameterised IQP circuits~\cite{recio2025train,armengol2025iqpopt} and is the foundation of the classical training procedure described in Sec.~\ref{sec:gen_learning}.

\begin{proposition}\label{prop:zd_classical}
    Given a tractable Fourier-phase circuit $U(\btheta)$ on $n$ qudits of dimension $d$, expectation values of Heisenberg-Weyl observables
    \begin{align}
        \langle\mathcal{D}(\k,\m)\rangle_{\btheta} = \bra{0}U^{\dagger}(\btheta)\,\mathcal{D}(\k,\m)\,U(\btheta)\ket{0}
    \end{align}
    can be estimated to additive error $\epsilon$ by a classical algorithm that uses $O(\operatorname{poly}(n,d)/\epsilon^2)$ time and
    $\operatorname{poly}(n,d)$ space. Furthermore, the estimator is unbiased. 
\end{proposition}
For a proof see App.~\ref{ap:hw-estimate}. One finds that the expectation value can be written
\begin{multline}\label{eq:hw_estimator}
    \langle\mathcal{D}(\k,\m)\rangle_{\btheta} =  \\ \mathbb{E}_{\z\sim \text{Unif}}\!\left[\exp\!\left(\frac{i\pi}{d}\,\m\cdot(2\z-\k)+i\Delta^{\k}_{\btheta}(\z)\right)\right],
\end{multline}
with
\begin{align}\label{eq:phase_diff}
    \Delta^{\k}_{\btheta}(\z) = \Phi_{\btheta}(\z) - \Phi_{\btheta}(\z\ominus\k),
\end{align}
the phase difference between $\z$ and $\z\ominus\k$, where $\z\ominus\k = \z - \k \mod d$. Note that since the term in square brackets in \eqref{eq:hw_estimator} is a unit modulus complex number, the expression is an expectation value of a bounded complex random variable with respect to the uniform distribution on $\z$. Chernoff-Hoeffding bounds then imply that the empirical estimate 
\begin{align}\label{eq:simexpval}
    \frac{1}{|\mathcal{Z}|}\sum_{\z\in\mathcal{Z}}\exp\!\left(\frac{i\pi}{d}\,\m\cdot(2\z-\k)+i\Delta^{\k}_{\btheta}(\z)\right),
\end{align}
formed by taking a batch $\mathcal{Z}=\{\z_i \sim \text{Unif}\}$ of uniformly sampled $\z$ vectors, is an unbiased estimate of $\langle\mathcal{D}(\k,\m)\rangle_{\btheta}$ with additive error $1/\sqrt{|\mathcal{Z}|}$. To achieve error $\epsilon$, one therefore requires $O(1/\epsilon^2)$ samples. Note that this is analogous to the $O(1/\epsilon^2)$ sample cost when estimating observable expectation values on quantum hardware, since observables are also estimated via an empirical mean of quantum circuit samples. 

If $\Phi$ is a smooth differentiable function of $\btheta$, it follows that gradients of a loss function 
\begin{align}
    L(\btheta) = f(\mathcal{D}(\k_1,\m_1), \cdots, \mathcal{D}(\k_s, \m_s))
\end{align}
where $f$ is also a smooth differentiable function can be estimated by implementing $\eqref{eq:simexpval}$ in software that implements automatic differentiation. The precision to which gradients are estimated is thus controlled by $|\mathcal{Z}|$ which adds a layer of stochasticity to the training.


\subsection{A Heisenberg-Weyl-generated phase layer}\label{sec:hw_phase}
We now construct a specific example of a tractable Fourier-phase unitary based on HW observables. In particular, we take $D(\btheta)$ to be generated by a subset of the operators $\mathcal{Q}(\g,\boldsymbol{0})$. We set
\begin{align}\label{eq:Dunitary_HW}
    D(\btheta) = \prod_{\g\in\mathcal{G}}\exp(i\theta_{\g}\mathcal{Q}_{\g}), \qquad \mathcal{Q}_{\g} = \mathcal{Q}(\g,\boldsymbol{0})
\end{align}
where $\mathcal{G}\subseteq\mathbb{Z}_d^n$ specifies the gate set and $\btheta = \{\theta_{\g}\}_{\g\in\mathcal{G}}$ collects the trainable parameters. The choice $\m = 0$ in the generators ensures that each $\mathcal{Q}_{\g}$ acts diagonally in the computational basis, so that all gates in the parameterised layer mutually commute and $D(\btheta)$ is indeed a diagonal phase unitary of the form \eqref{eq:Dunitary}. Using \eqref{eq:Q_def} together with $Z^{g}\ket{z} = \omega^{gz}\ket{z}$, each gate generator acts on $\ket{\z}$ as
\begin{align}
    \mathcal{Q}_{\g}\ket{\z} &= \prod_{j=1}^n\big(\chi\,\omega^{g_j z_j} + \chi^*\,\omega^{-g_j z_j}\big)\ket{\z} \\ \label{eq:phigz}
    &= \prod_{j=1}^n\sqrt{2}\cos\!\left(\frac{2\pi z_j g_j}{d}+\frac{\pi}{4}\right)\ket{\z}  \\
    &= \phi_{\g}(\z)\ket{\z}.
\end{align}
The phase function $\Phi_{\btheta}$ entering \eqref{eq:Dunitary} therefore takes the explicit form
\begin{align}\label{eq:phi_HW}
    \Phi_{\btheta}(\z) = \sum_{\g\in\mathcal{G}}\theta_{\g}\,\phi_{\g}(\z),
\end{align}
and can be computed efficiently by an algorithm in $O(|\mathcal{G}|n)$ time. Since $O(|\mathcal{G}|)=O(\text{poly}(n))$ by assumption of the circuit being efficient on quantum hardware, we can compute $\Phi_{\btheta}(\z)$ to machine precision by a classically efficient algorithm. 

In general, since $\mathcal{G}\subseteq
\mathbb{Z}_d^n$, there is an exponentially large set of possible generators to choose from. To build practical models we therefore need to work with restricted gate sets. Two natural restrictions are to limit the \emph{weight} of $\g$, defined as the number of nonzero entries (which corresponds to a $k$-body interaction), and to limit the support of each $g_i$ to a subset of $\mathbb{Z}_d$ of size less than $d$. We will return to this point in Sec.~\ref{sec:gen_learning}, where we explain how both these restrictions are natural for generative learning settings.

\subsection{Efficient batch estimation on hardware}\label{sec:efficient_software}
Although \eqref{eq:hw_estimator} shows that expectation values of HW observables can be estimated efficiently, in order to efficiently train large quantum generative models we will need the ability to estimate batches of such observables in parallel. Here we describe how this can be achieved via linear algebra, which leads to a particularly efficient batch estimation procedure.

Concretely, suppose we wish to estimate a batch of $|\mathcal{O}|$ expectation values
\begin{align}
    \mathcal{O} = \{ \mathcal{O}_i\}_{i=1}^{|\mathcal{O}|} = \big\{\langle\mathcal{D}(\k_i,\m_i)\rangle_{\btheta}\big\}_{i=1}^{|\mathcal{O}|}.
\end{align}
We show in App.~\ref{ap:batching} that the $|\mathcal{O}|$ expectation values can be estimated as
\begin{align}\label{eq:hw_batch}
    (\mathcal{O}_1, \cdots, \mathcal{O}_{|\mathcal{O}|})\approx \frac{1}{|\mathcal{Z}|}\exp(i (J + E))\mathbf{1}_{|\mathcal{Z}|},
\end{align}
where $\exp(\cdot)$ denotes \emph{element-wise} application of the exponential function and $\mathbf{1}_{|\mathcal{Z}|}$ is a $|\mathcal{Z}|\times 1$ matrix of ones. The $|\mathcal{O}|\times|\mathcal{Z}|$ matrix $J$ is given by  
\begin{align}\label{eq:Jmatrix}
    J = \frac{\pi}{d}\big(2 M Z^T - ((M\odot K)\mathbf{1}_n)\,\boldsymbol{1}_{|\mathcal{Z}|}^T\big)\,
\end{align}
where $K,M\in\mathbb{Z}_d^{|\mathcal{O}|\times n}$ stack the $\k$ and $\m$ vectors row-wise and $Z\in\mathbb{Z}_d^{|\mathcal{Z}|\times n}$ stacks the samples $\z$ row-wise. The matrix $E$ is given by 
\begin{align}
    E &= \sum_{w=1}^{w_{\max}}E_w
\end{align}
where $w_{\max}$ is the largest weight of any gate, that is, the maximum number of qudits on which it acts non-trivially. The $E_w$ are given by
\begin{multline}\label{eq:factored_E}
    E_{w} = \sum_{\boldsymbol{\sigma}\in\{0,1\}^{w}}(-1)^{|\boldsymbol{\sigma}|+1}\,(\mathbf{C}^{w}_{\boldsymbol{\sigma}})^T\,\mathrm{diag}(\boldsymbol{\theta}_{w})\,\mathbf{B}^{w}_{\boldsymbol{\sigma}}\; \\ +\;\boldsymbol{1}_{|\mathcal{O}|}\boldsymbol{\theta}_{w}^T\widetilde{\mathbf{B}}_{w},
\end{multline}
where the matrices $\mathbf{B}^{w}_{\boldsymbol{\sigma}}, \widetilde{\mathbf{B}}_{w}, \mathbf{C}^{w}_{\boldsymbol{\sigma}}$ (defined in App.~\ref{ap:batching}) have shape $|\mathcal{G}_w|\times |\mathcal{Z}|$, $|\mathcal{G}_w|\times |\mathcal{Z}|$ and $|\mathcal{G}_w|\times |\mathcal{O}|$ respectively, and where $\mathcal{G}_w$ contains all gates that act non-trivially on exactly $w$ qudits and $\btheta_w$ are the corresponding parameters. The runtime complexity of estimating the $|\mathcal{O}|$ expectation values is thus 
\begin{align}\nonumber
    O\!\,\Big(|\mathcal{O}|\,|\mathcal{Z}|\,n \;+\; \sum_{w} 2^w\,|\mathcal{G}_w|\big((|\mathcal{O}|+|\mathcal{Z}|) + |\mathcal{O}||\mathcal{Z}|\big)\Big).
\end{align}
Note that as we discuss in Sec.~\ref{sec:spectralbias}, one may often wish to keep $w_{\text{max}}$ reasonably small (for example one- and two-qudit gates), and thus, although the number of terms appearing in the sum grows exponentially with $w$, there will often be relatively few terms to compute. For example, the computation of $E$ is roughly 6 times more expensive than the case for parameterized IQP circuits when the models contain the same number of one and two-qudit gates.

\section{Spectral Born machines}\label{sec:gen_learning}

We are now ready to assemble the ingredients of the previous sections into a generative-learning pipeline for distributions over $\mathbb{Z}_d^n$. Following the same setup as the qubit IQP case~\cite{recio2025train,rudolph2024trainability,liu2018differentiable}, we define a class of generative models $ q_{\btheta}(\x)$ over $\mathbb{Z}_d^n$, that we call \emph{spectral Born machines}, via the computational measurement distribution of the fiducial state $\ket{0}$ evolved under a Fourier-phase circuit $U(\btheta)$:
%
\begin{align}\nonumber
    q_{\btheta}(\x)
    = \operatorname{Tr}\!\left[
    \left(\ket{\x}\!\bra{\x}\otimes \mathbb{I}\right)\rho_{\btheta}
    \right],
    \quad
    \rho_{\btheta} = U(\btheta)\ket{0}\!\bra{0}U^{\dagger}(\btheta).
\end{align}
Note we allow for the possibility that the distribution is a marginal distribution of the total circuit, since it is known this can be beneficial to increase expressivity in IQP Born machines \cite{kurkin2025universality}. Given a dataset $\mathcal{X}=\{\x_i\}$ whose elements are sampled $\mathrm{iid}$ from a ground truth distribution $p$, the goal is to infer parameters $\btheta$ such that $q_{\btheta}$ is close to $p$ in some natural sense. The precise notion of closeness can vary \cite{theis2015note}, however here we will focus on the $\mathrm{MMD}$ distance, which is a common evaluation metric used in the classical literature \cite{xu2018empirical,bischoff2024practical} and has the benefit that it can be classically estimated on trained spectral Born machines. 

For the case $d=2$, spectral Born machines become the class of IQP Born machines and thus inherit similar complexity-theoretic classical hardness guarantees related to sampling \cite{recio2025train, marshall2024improved, bremner2011classical,bremner2016average}. This makes a general dequantization of sampling $q_{\btheta}(\x)$ unlikely, and motivates why such a model class may have advantages relative to classical generative models. However, we acknowledge this is not a bullet-proof guarantee against dequantization in typical learning scenarios, where trained parameters are likely to be relatively small and distributions spectrally concentrated, and more work is needed to understand the classical complexity of the model under such constraints.   

\subsection{Spectral bias of the model class}\label{sec:spectralbias}
One potentially powerful property of spectral Born machines relates to a spectral bias that is a consequence of their $\mathbb{Z}_d^n$ Fourier structure. Let
\begin{align}
    \hat{\psi}_{\btheta}(\z) := \frac{1}{\sqrt{d^n}}e^{i\Phi_{\btheta}(\z)}
\end{align}
denote the Fourier-basis amplitudes of the state immediately after the diagonal phase layer. Since the final layer is an inverse Fourier transform, these are precisely the Fourier coefficients of the output state amplitudes. For the output distribution $q_{\btheta}$ its Fourier coefficients are
\begin{align}\label{eq:model_autocorr}
    \hat q_{\btheta}(\k)
    &= \langle \mathcal{D}(\k,\boldsymbol{0})\rangle_{\btheta} \nonumber \\
    &= \frac{1}{d^n}\sum_{\z\in\mathbb{Z}_d^n}
    \exp\!\left(i\Phi_{\btheta}(\z)-i\Phi_{\btheta}(\z\ominus\k)\right) \nonumber \\
    &= \sum_{\z\in\mathbb{Z}_d^n}
    \hat{\psi}_{\btheta}(\z)\,
    \hat{\psi}_{\btheta}^{*}(\z\ominus\k).
\end{align}
Thus the Fourier coefficients of the probability distribution are the circular autocorrelation of the Fourier coefficients of the quantum state.

This gives a simple way to understand the smoothness bias induced by restricting the gate set $\mathcal{G}$. If the phase layer is built from low-order generators, then from \eqref{eq:phigz} $\Phi_{\btheta}$ is itself a low-frequency function on $\mathbb{Z}_d^n$: it varies only through low-order Fourier modes, or through low-weight interactions between sites. The exponential $e^{i\Phi_{\btheta}}$ can of course mix frequencies, so this is not an exact band-limit in general. Nevertheless, for moderately small parameter values, the resulting Fourier-basis state remains naturally biased toward low-order spectral structure. This allows us to bias the model distribution towards smoothness along two natural axes: (i) by limiting the maximum degree of any $g_i$ appearing in a gate generator we bias towards smoothness with respect to cyclically permuting the labels, and (ii) by limiting the maximum weight of any $\g$ we bias towards distributions whose statistics are captured by `low body' correlations. Given the importance of smoothness for generalization \cite{rosca2020case, dherin2022neural} and the fact that the success of neural networks may be due to similar spectral biases \cite{rahaman2019spectral,xu2019frequency}, we believe this is a property of the model class that deserves closer attention. 

\subsection{Loss functions via graph-Fourier kernels}\label{sec:graph_mmd}
To train the model we minimise a maximum mean discrepancy loss \cite{gretton2012kernel,li2017mmd,li2017mmd,rudolph2024trainability,liu2018differentiable} between $q_{\btheta}$ and the empirical distribution of training data $\mathcal{X} = \{\x_1,\ldots,\x_N\}\subset\mathbb{Z}_d^n$. Unlike the case of binary data, where the natural notion of closeness between data points is the Hamming distance, the `correct' notion of closeness in $\mathbb{Z}_d^n$ will generally depend on the specific type of discrete data. A natural way to formalize this is to define a weighted graph $G_i = (V_i, E_i, w_i)$ for each variable, where the notion of closeness between the labels $u,v \in \mathbb{Z}_d$ for element $i$ is given by a symmetric edge weight $w_i(u,v) \ge 0$, with larger values implying closer labels. Closeness across sites is then captured by the Cartesian product graph $G = G_1\,\square\,\cdots\,\square\,G_n$ on $\mathbb{Z}_d^n$. 

In the following we explain how to build MMD loss functions based on this notion of closeness via graph spectral analysis \cite{kondor2002diffusion, shuman2013emerging, smola2003kernels}. We will then focus specifically on shift invariant graphs, since in this case the resulting MMD diagonalises into a probabilistic mixture of squared differences of Heisenberg-Weyl moments. This naturally leads us to two types of discrete data, stemming from cyclic and complete graphs:
\begin{itemize}
    \item \emph{Cycle graphs}: Neighboring labels (mod $d$) are close to each other. This relates to cyclic, ordinal data such as clock times, days of the week, compass directions and colours.
    \item \emph{Complete graphs}: All labels are equally close. This relates to categorical data without an obvious order. For example fruit, countries and words. 
\end{itemize}
%



To construct the MMD loss we first consider the graph Laplacian of $G_i$, defined as $L_i = D_i - A_i$, where $A_i$ is the weighted adjacency matrix with entries $[A_i]_{uv} = w_i(u,v)$ and $D_i$ is the diagonal matrix of weighted vertex degrees, $[D_i]_{uu} = \sum_v w_i(u,v)$. Non-negativity of the weights guarantees that $L_i$ is positive semi-definite and so can be diagonalised,
\begin{align}
L_i = U_i\Lambda_i U_i^{\dagger}= \sum_k \lambda_k^{(i)} \u_k^{(i)}(\u_k^{(i)})^{\dagger}
\end{align}
with eigenvalues $0=\lambda_0^{(i)}\le\cdots\le\lambda_{d-1}^{(i)}$, and orthonormal eigenvectors $\u_k^{(i)}\in\mathbb{C}^d$. The Cartesian product graph $G = G_1\,\square\,\cdots\,\square\,G_n$ has corresponding Laplacian
\begin{align}
    L = \sum_{i=1}^n I^{\otimes(i-1)}\otimes L_i\otimes I^{\otimes(n-i)}
\end{align}
with eigenvectors and eigenvalues
\begin{align}\label{eq:lap_eig}
    \u_{\k} = \bigotimes_i \u_{k_i}^{(i)}, \qquad \Lambda_{\k} = \sum_{i=1}^n \lambda_{k_i}^{(i)}, \qquad \k = (k_1,\ldots,k_n).
\end{align}
A \emph{spectral kernel} is any matrix 
\begin{align}
    \mathsf{K} = f(L) \qquad f:\mathbb{R}_{\ge 0}\to\mathbb{R}_{\ge 0},
\end{align}
that is, it applies a function $f$ to the eigenvalues of $L$. We define the corresponding spectral kernel function
\begin{align}\label{eq:spectral_kernel}
    \kappa(\x,\y) = \mathbf{e}_{\x}^{\top}\,\mathsf{K}\,\mathbf{e}_{\y},
\end{align}
where $\mathbf{e}_{\x}=\mathbf{e}_{x_1}\otimes \cdots \otimes \mathbf{e}_{x_n}\in\mathbb{R}^{d^n}$ denotes the standard-basis vector for vertex $\x\in\mathbb{Z}_d^n$, so that $\kappa(\x,\y)$ is the $(\x,\y)$ entry of $\mathsf{K}$.  The squared MMD between distributions $p$ and $q$ is then defined in the usual way as
\begin{multline}\label{eq:mmd_def}
    \mathrm{MMD}^2(p,q) = \mathbb{E}_{\x,\x'\sim p}[\kappa(\x,\x')]  - 2\,\mathbb{E}_{\x\sim p,\,\y\sim q}[\kappa(\x,\y)] \\ + \mathbb{E}_{\y,\y'\sim q}[\kappa(\y,\y')],
\end{multline}
which corresponds to the squared distance in the reproducing kernel Hilbert space induced by $\kappa$
between the expected feature means of $p$ and $q$ \cite{gretton2012kernel}.

Defining the graph-Laplacian moments of a distribution $p$ on $\mathbb{Z}_d^n$ as
\begin{align}\label{eq:graph_moment}
    \tilde p(\k) := \mathbb{E}_{\x\sim p}\!\big[\mathbf{e}_{\x}^{\top}\u_{\k}\big],
\end{align}
the squared MMD with kernel $\kappa$ can be written
\begin{align}\label{eq:mmd_graph}
    \mathrm{MMD}^2(p,q) = \sum_{\k}\hat\kappa(\k)\big|\tilde p(\k) - \tilde q(\k)\big|^2,
\end{align}
a non-negatively weighted $\ell^2$ distance between moments of $p$ and $q$, with weights $\hat\kappa(\k)=f(\Lambda_{\k})$ given by the eigenvalues of the kernel matrix $\mathsf{K}$. 

For concreteness we focus on the heat (diffusion) kernel $f(\lambda) = e^{-t\lambda}$ throughout, which is the canonical choice and depends on a single positive bandwidth parameter $t > 0$. This choice is natural for machine learning applications since it allows us to target smooth functions: for a function $h:\mathbb{Z}_d^n\rightarrow \mathbb{R}$ written as a vector $\h\in\mathbb{R}^{d^n}$, a natural notion of smoothness with respect to the graph structure is the graph Dirichlet energy,
\begin{align}\label{eq:dirichlet}
\h^{\dagger}L\h=\frac{1}{2}\sum_{u,v}w(u,v)|h(u)-h(v)|^2=\sum_{\k}\Lambda_{\k}|\tilde h_{\k}|^2    
\end{align} 
where $\h= \sum_{\k}\tilde{h}_{\k}\u_{\k}$ is the decomposition of $\h$ in terms of Laplacian eigenvectors. This can be seen as a measure of smoothness since in order to lower the value of the central expression of \eqref{eq:dirichlet}, one should prioritize making function values of close variables $u$, $v$ similar, since these correspond to larger weights $w$. Components $\tilde{h}_{\k}$ with small $\Lambda_{\k}$ are thus smooth modes that contribute little to the energy. In the same way that the qubit MMD acts as a low-pass filter to target smooth modes of $\mathbb{Z}_2^n$, the choice $f(\lambda) = e^{-t\lambda}$ acts as a low-pass filter to target smooth modes of $\mathbb{Z}_d^n$ with respect to the closeness defined by the graph structure. We now consider two specific graph structures that are particularly relevant for this work. 

\subsubsection{Shift-invariant graphs: cycle $C_d$ and complete $K_d$}
When each $G_i$ is invariant under cyclic shifts of labels, which is the case for the cycle graph or the complete graph, the Laplacian $L_i$ is a circulant matrix and the eigenvectors $\u_{\k}$ of the Laplacian are the characters of $\mathbb{Z}_d^n$,
\begin{align}\label{eq:char_zdn}
    \mathbf{e}_{\x}^{\top}\u_{\k} = \frac{1}{\sqrt{d^n}}\,\omega^{\k\cdot\x}.
\end{align}
Since the kernel $\mathsf{K} = f(L)$ is also diagonalised by this basis it is circulant on $\mathbb{Z}_d^n$ and so the kernel function is translation invariant, $\kappa(\x,\y) = \kappa(\x\ominus\y)$. Substituting \eqref{eq:char_zdn} into the graph-Fourier moment \eqref{eq:graph_moment} gives
\begin{align}\label{eq:moment_to_obs}
    \tilde p(\k) = \frac{1}{\sqrt{d^n}}\,\mathbb{E}_{\x\sim p}\!\big[\omega^{\k\cdot\x}\big] = \frac{1}{\sqrt{d^n}}\,\langle\mathcal{D}(\k,0)\rangle_p.
\end{align}
Substituting into \eqref{eq:mmd_graph}, the loss takes the form
\begin{align}\label{eq:zd_mmd_observable}
    \mathrm{MMD}^2(p,q) &= \frac{1}{d^n}\sum_{\k\in\mathbb{Z}_d^n} \hat\kappa(\k) \big|\langle\mathcal{D}(\k,0)\rangle_p - \langle\mathcal{D}(\k,0)\rangle_q\big|^2, \nonumber \\
    &= \mathbb{E}_{\k \sim \text{Unif}}\left[\hat\kappa(\k) \big|\langle\mathcal{D}(\k,0)\rangle_p - \langle\mathcal{D}(\k,0)\rangle_q\big|^2\right], \nonumber 
\end{align}
i.e.\ a non-negatively weighted mixture of squared differences of the Heisenberg-Weyl Fourier moments. Since rescaling the kernel by an overall positive constant rescales the loss but leaves all training trajectories unchanged, we are free to absorb $\sum_{\k}\hat\kappa(\k)$ into the normalisation. We thus define the probability mass function
\begin{align}
    \mathcal{P}_{t}(\k) = \frac{\exp(-t \Lambda_{\k})}{\sum_{\k'}\exp(-t \Lambda_{\k'})},
\end{align}
and work with the rescaled MMD loss function:
\begin{align}\label{eq:zd_mmd_mixture}
    \mathrm{MMD}^2(p,q) \;=\; \mathbb{E}_{\k\sim\mathcal{P}_t}\!\Big[\,\big|\langle\mathcal{D}(\k,0)\rangle_p - \langle\mathcal{D}(\k,0)\rangle_q\big|^2\,\Big].
\end{align}

\begin{figure*}
    \centering
    \includegraphics[width=\textwidth]{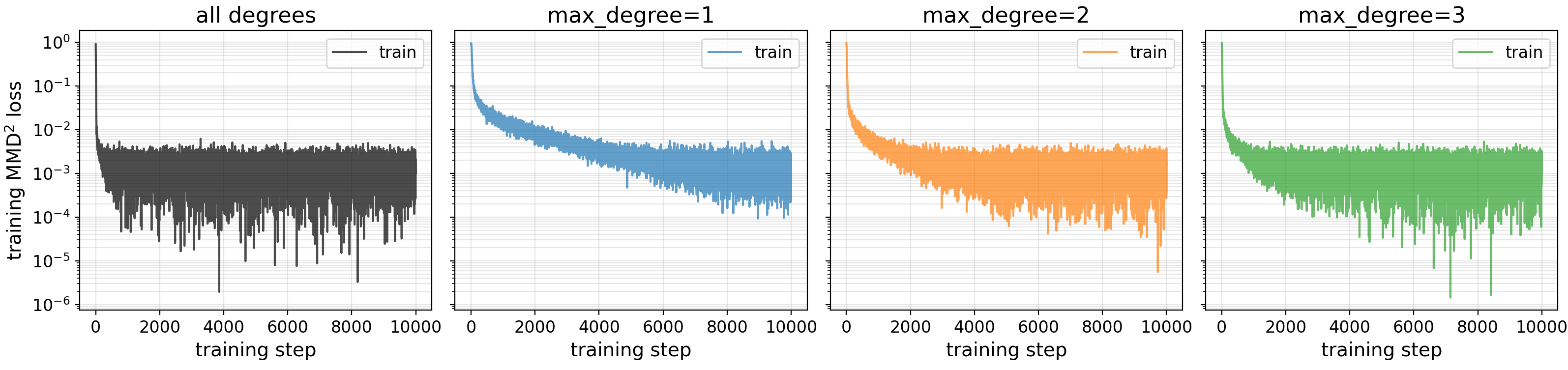}
    \caption{The four spectral Born machines (with varying parameter counts) used for Potts model experiments achieve similar training loss, suggesting that the models converge towards similar test performance regardless of model complexity.}
    \label{fig:potts_train}
\end{figure*}

Since we are working with product graphs, from  \eqref{eq:lap_eig} we have $\mathcal{P}_t(\k) = \prod_i\mathcal{P}_t(k_i)$. The choice of each $\mathcal{P}_t(k_i)$ should depend on whether the element $i$ of the data is of (cyclic) ordinal type or categorical. For the former case, the cycle graph gives us 
\begin{align}\label{eq:cycle_marginal}
    \mathcal{P}_t(k_i) = \frac{e^{-4t\sin^2(\pi k_i/d)}}{Z}
\end{align}
whereas for categorical variables the complete graph gives
\begin{align}\label{eq:kd_marginal}
    \mathcal{P}_t(k) = \begin{cases}\pi_0 & k = 0,\\[2pt] (1-\pi_0)/(d-1) & k\ne 0,\end{cases} 
\end{align}
with 
\begin{align}
    \pi_0 = \frac{1}{1+(d-1)\,e^{-td}}.
\end{align}
The result is expected given the Fourier structure. For cyclic ordinal variables, smoothness is naturally tied to low order Fourier coefficients, which have a higher chance of being sampled. Since categorical variables do not have any ordering, all cyclic shifts of labels are equally important, and all nonzero Fourier coefficients are sampled with equal probability.

\subsubsection{Finite-sample estimation}
Here we show how to efficiently estimate and train the $\mathrm{MMD}^2$ loss. Define $p_{\mathcal{X}}$ to be an empirical distribution of the dataset $\mathcal{X}$ sampled from the ground truth distribution $p$
\begin{align}
    p_{\mathcal{X}}(\x) = \frac{1}{|\mathcal{X}|}\sum_{\y\in\mathcal{X}}\mathbb{I}[\y=\x],
\end{align}
with $\mathbb{I}$ the indicator function. Note that $\langle\mathcal{D}(\k,0)\rangle_{p_{\mathcal{X}}}$ can be computed exactly and efficiently since $p_{\mathcal{X}}$ has sparse support. Let $\widehat{\langle\mathcal{D}(\k,\m)\rangle}_{\btheta}$ be the unbiased estimators of $\langle\mathcal{D}(\k,\m)\rangle_{\btheta}$ obtained via the batched estimator of Sec.~\ref{sec:efficient_software}. Since the $\mathrm{MMD}^2(p,q)$ is an expectation value with respect to sampling $\k$, we may estimate it by sampling a set $\mathcal{K}=\{\k_i\sim \mathcal{P}_t\}_{i=1}^{|\mathcal{K}|}$ and using the estimator
\begin{align}\nonumber
    \widehat{\mathrm{MMD}^2}(\mathcal{X},\mathcal{K},\mathcal{Z},\btheta) = \frac{1}{|\mathcal{K}|}\sum_{\k\in\mathcal{K}}  |\langle\mathcal{D}(\k,0)\rangle_{p_{\mathcal{X}}} - \widehat{\langle\mathcal{D}(\k,0)\rangle}_{\btheta}\big|^2.
\end{align}
\begin{figure}
    \centering
    \includegraphics[width=0.5\linewidth]{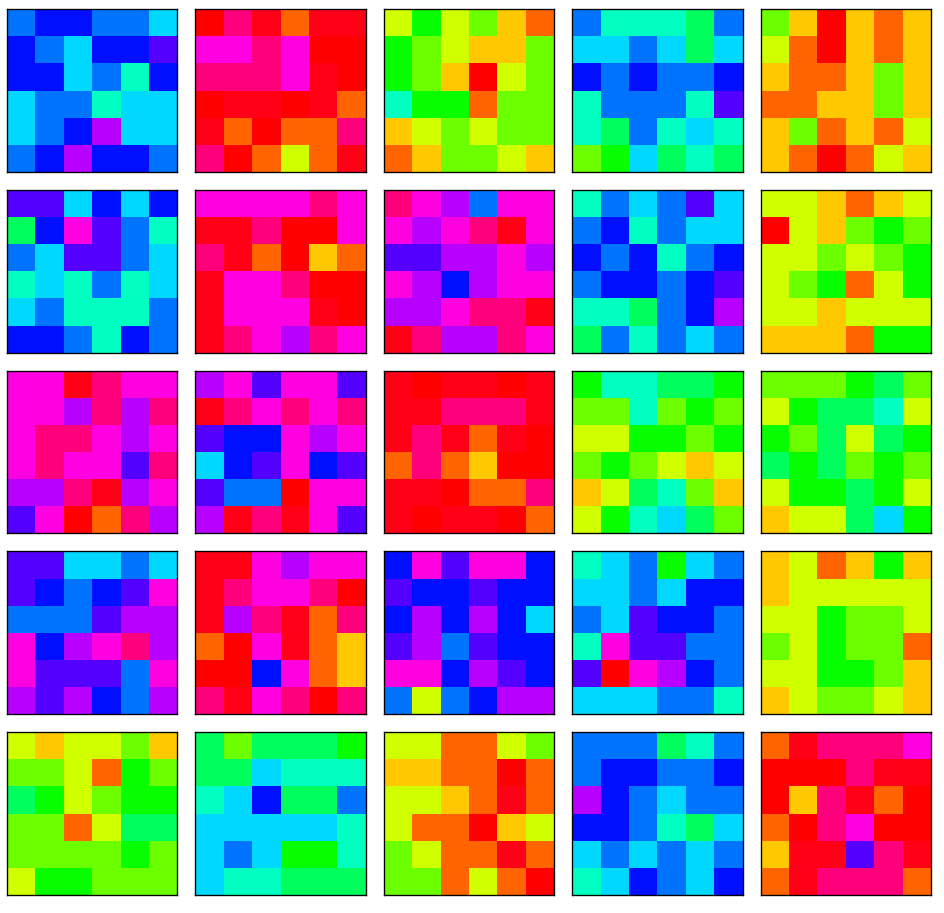}
    \caption{Example configurations from the Potts model on a $6\times 6$ lattice. Colours represent each of the 16 possible values the variables can take.}
    \label{fig:pottsegs}
\end{figure}
This estimator will be biased however due to the quadratic non-linearity. In App.~\ref{ap:mmd-estimation} we show how to convert this into an unbiased estimator $\widehat{\mathrm{MMD}_u^2}(\mathcal{X},\mathcal{K},\mathcal{Z},\btheta)$, in the sense that 
\begin{align}\nonumber
    \mathbb{E}_{\mathcal{X}\sim p, \mathcal{K}\sim \mathcal{P}_t,\mathcal{Z}\sim \text{Unif}}\left[ \widehat{\mathrm{MMD}_u^2}[\mathcal{X},\mathcal{K},\mathcal{Z},\btheta] \right] = \mathrm{MMD}^2(p,q_{\btheta}).
\end{align}
Since $ \widehat{\mathrm{MMD}^2}(\mathcal{X},\mathcal{K},\mathcal{Z},\btheta)$ is a smooth differentiable function of $\btheta$, we can use automatic differentiation to obtain unbiased gradients of $\mathrm{MMD}^2$ in the same fashion as IQP Born machines, and use these to train the model via gradient descent. 

\section{Numerical experiments}
In this section we present two demonstrations of training spectral Born machines and investigate their properties. We train the models via the \texttt{tcdq} module of PennyLane using an H200 Nvidia GPU. 

\begin{figure*}
    \centering
    \includegraphics[scale=0.5]{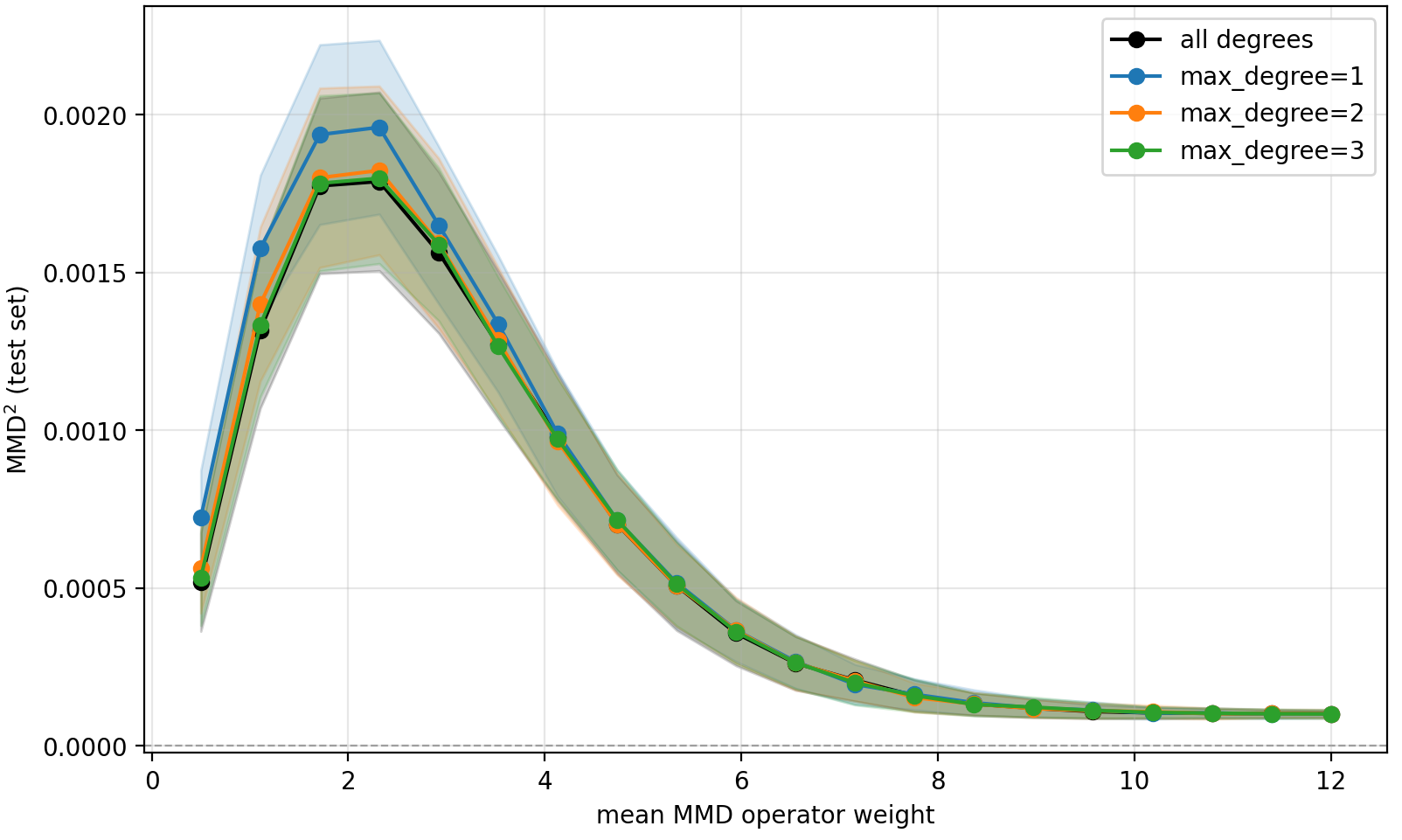}
    \caption{The inductive bias of the spectral Born machine allows small models to achieve comparable performance compared to significantly larger models. Curves show test loss values for the four models used in the Potts experiment as a function of the mean operator weight of the complete graph $\mathrm{MMD}$ loss (shaded areas denote 1 standard deviation of the estimated $\mathrm{MMD}^2$).}
    \label{fig:potts:eval}
\end{figure*}

\subsection{Learning the Potts model}\label{sec:potts}
In order to probe the integer inductive bias described in Sec.~\ref{sec:spectralbias} we consider a synthetic dataset whose ground-truth distribution is genuinely cyclic and known by construction: the $d$-state clock model, a vector Potts model on $\mathbb{Z}_d$ \cite{wu1982potts}. In this model each site $i$ carries a label $x_i\in\mathbb{Z}_d$, interpreted as a `clock hand' pointing at the angle $2\pi x_i/d$. Sites interact ferromagnetically through the cyclic coupling
\begin{align}\label{eq:clock_energy}
    E(\x) = -J\sum_{(i,j)\in\mathcal{E}}\cos\!\left(\frac{2\pi}{d}(x_i - x_j)\right),
\end{align}
where $\mathcal{E}$ is the edge set of an interaction graph and $J>0$ favours alignment of neighboring hands. The data distribution is the Boltzmann distribution $p(\x)\propto e^{-E(\x)/T}$ at temperature $T$. Due to the form of \eqref{eq:clock_energy}, small cyclic shifts in labels relate to small changes in energy and so $p$ is smooth with respect to the cyclic distance on $\mathbb{Z}_d$. This makes the distribution a natural tool to test the $\mathbb{Z}_d$ inductive bias of our generative models. 

We take $d=16$ and place the sites on a $6\times 6$ periodic square lattice ($n=36$ sites), so that the target is a distribution over $\mathbb{Z}_{16}^{36}$, encodable in a quantum circuit of $36\log_2 16 = 144$ qubits. We fix $J=1$ and sample at temperature $T=0.7$. 
Samples are generated by Metropolis Monte Carlo and split into a training and test sets of size $10^4$ each. Example configurations generated in this way are shown in Fig.~\ref{fig:pottsegs}. 

To learn this distribution we train four different spectral Born machines with $36$ qudits of dimension $d=16$. The models differ by the chosen gate set. Each model is constructed using all one-qudit gates and all two-qudit gates with the following restrictions on the possible values the $g_i$ can take in the gate generators $\g$: 
\begin{itemize}
\item \emph{Degree 1}: $g_i\in\{15,0,1\}$ (2592 parameters)
\item \emph{Degree 2}:  $g_i\in\{14,15,0,1,2\}$ (10224 parameters)
\item \emph{Degree 3}: $g_i\in\{13,14,15,0,1,2,3\}$ (22896 parameters)
\item \emph{All degrees}: $g_i\in\{0,\cdots,15\}$ (142290 parameters)
\end{itemize}
Since small and large Fourier coefficients correspond to small clockwise or anticlockwise shifts in direct space, each of the first three choices biases the model towards smoothness on $\mathbb{Z}_d^n$ with increasing expressivity. The final choice is not biased in this way since it contains all degrees.

\begin{figure*}
    \centering
    \includegraphics[width=\textwidth]{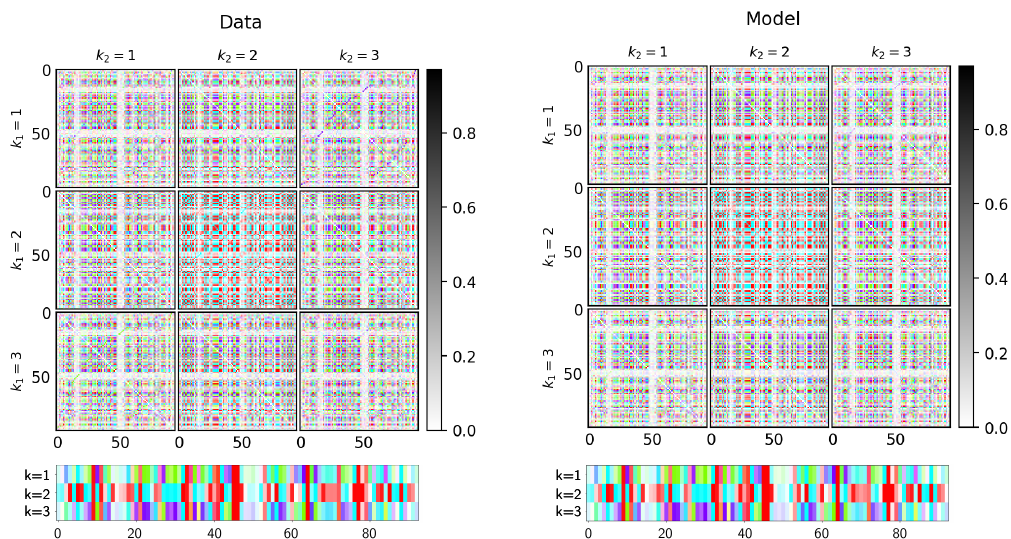}
    \caption{Comparing visualizations of the Fourier coefficients of the complete dataset (left) versus trained spectral Born machine (right) suggests that the models learn broadly similar features to the data. The panels below show the values of the Fourier coefficients with vectors $\k$ of weight 2, where the $(i,j)^{th}$ elements of $\k$ take one of the nine possible values $(k_i,k_j)$. For instance, the first row of the panel below shows the first order Fourier coefficients (diagonal elements) of the $k_1=k_2=1$ subplot of the figure above. Since coefficients are complex, the hue shows the magnitude (with scale given to the right of the plots) and the colour shows the phase.}
    \label{fig:rna_plots}
\end{figure*}

All models are trained with $|\mathcal{K}|=|\mathcal{Z}|=1000$ for 10000 iterations with an initial learning rate of $0.001$. To train we use the $\textrm{MMD}^2$ loss based on the complete graph, with a decay parameter $t$ such that the average weight of operators sampled via \eqref{eq:kd_marginal} is 3. We choose the complete graph in order to test the inductive bias of the model, not the loss signal: in this way we do not bias the training to favor the cyclically-biased gate sets. Single qudit parameters are initialized in such a way that, if all other parameters are set to zero, the initial distribution is the uniform distribution. Two-qudit gate parameters are initialized around zero using a normal distribution with standard deviation $0.001$. 

From Fig.~\ref{fig:potts_train} we see that the four models reach similar final loss values, suggesting similar test performance. This is confirmed in Fig.~\ref{fig:potts:eval}, which shows similar squared MMD distances to the test set for different values of the decay parameter $t$ (plotted in terms of the corresponding average \textrm{MMD} operator weight). For the value of $t$ used for training (mean weight 3), all models achieve a test $\textrm{MMD}^2\approx 0.0016$, which closely matches the loss values achieved during training and therefore does not suggest overfitting. Note that, due to finite data, the standard error of the Fourier coefficients in the empirical training distribution is on the order of $1/\sqrt{|\mathcal{X}|}=0.01$. A $\textrm{MMD}^2$ of $0.0016$ suggests the Fourier coefficients of the trained model have average errors close to $\sqrt{0.0016}\approx 0.04$, which, being close to $0.01$, implies the Fourier structure has been learned reasonably close to the inherent statistical uncertainty present in the data. 

We see that the lack of expressivity of the degree-1 model leads to slightly worse performance, however all other models perform essentially the same. This suggests that the inductive bias of the model is playing the desired role, since it allows us to parameterize only the relevant transformations that are required for learning, leading to equal performance with far fewer parameters. Interestingly, the largest model converges in the smallest number of iterations, which may be due to the advantages of overparameterization. We remark that this does not necessarily imply more efficient learning, since gradient estimation steps are significantly slower due to the increased parameter count. 


\subsection{Learning of RNA sequences}\label{sec:dna}
Here we train the model on real data by learning the distribution of ribosomal RNA (Rfam RF00001 \cite{griffiths2003rfam}), a highly conserved non-coding RNA molecule that serves as an important structural and functional component of the large ribosomal subunit across all domains of life. The Rfam RF00001 data has a length of 120 nucleotides, however some entries contain gaps in the RNA sequence due to evolutionary insertions and deletions. To obtain a dense dataset without gaps, we use a filtered representation for model training. Starting from the Stockholm seed alignment, we retain only alignment columns with low gap frequency and unambiguous nucleotide identities, then discard sequences that still contain gaps in the retained columns. This produces fixed-length dense vectors over the four nucleotide states A, C, G, and T. Under this pre-processing, the dataset has 93 retained positions rather than the full 120. We split the data into a training set of size 428 and a test set of size 53. This is a small amount of data, which effectively puts us in a scarce data regime where deep neural networks are susceptible to overfitting. 

To train the model we use a Spectral Born machine with 95 qudits ($d=4$) that parameterizes a distribution over $\mathbb{Z}_4^{93}$, via the marginal distribution of the first $93$ qudits. By encoding to qubits, this results in a quantum circuit of 190 qubits. We treat the symbols $A,C,G,U$ as categorical variables, using the corresponding complete graph $\mathrm{MMD}^2$ loss with a value $t$ chosen to target two-qudit correlations on average. We consider two different models corresponding to different choices of gate set:
\begin{enumerate}
    \item \emph{All two-qudit gates}: Every possible two-qudit phase gate is used (40470 parameters)
    \item \emph{two and three-qudit gates}: All two qudit gates plus a selection of three qudit gates (1,297,661 parameters)
\end{enumerate}
In the second choice, we compute the values of the empirical Fourier coefficients of the training data, sort them by magnitude, then select the gates (acting only on the data qudits, not ancillas) that correspond to the largest coefficients until the parameter count is reached. 

The final training loss value of $\mathrm{MMD}^2\approx 0.0020, 0.0035$ suggests the low order model and empirical Fourier coefficients differ on average by approximately $\sqrt{0.0020}=0.04$ and $\sqrt{0.0035}\approx0.06$ (assuming errors are roughly uniform across low order coefficients). As with the Potts model, this is in line with the inherent statistical uncertainty in the training data which is on the order of $1/\sqrt{|\mathcal{X}|}\approx 0.05$. This therefore does not obviously indicate severe overfitting, despite the large parameter count. 

\begin{figure}
    \centering
    \includegraphics[width=\linewidth]{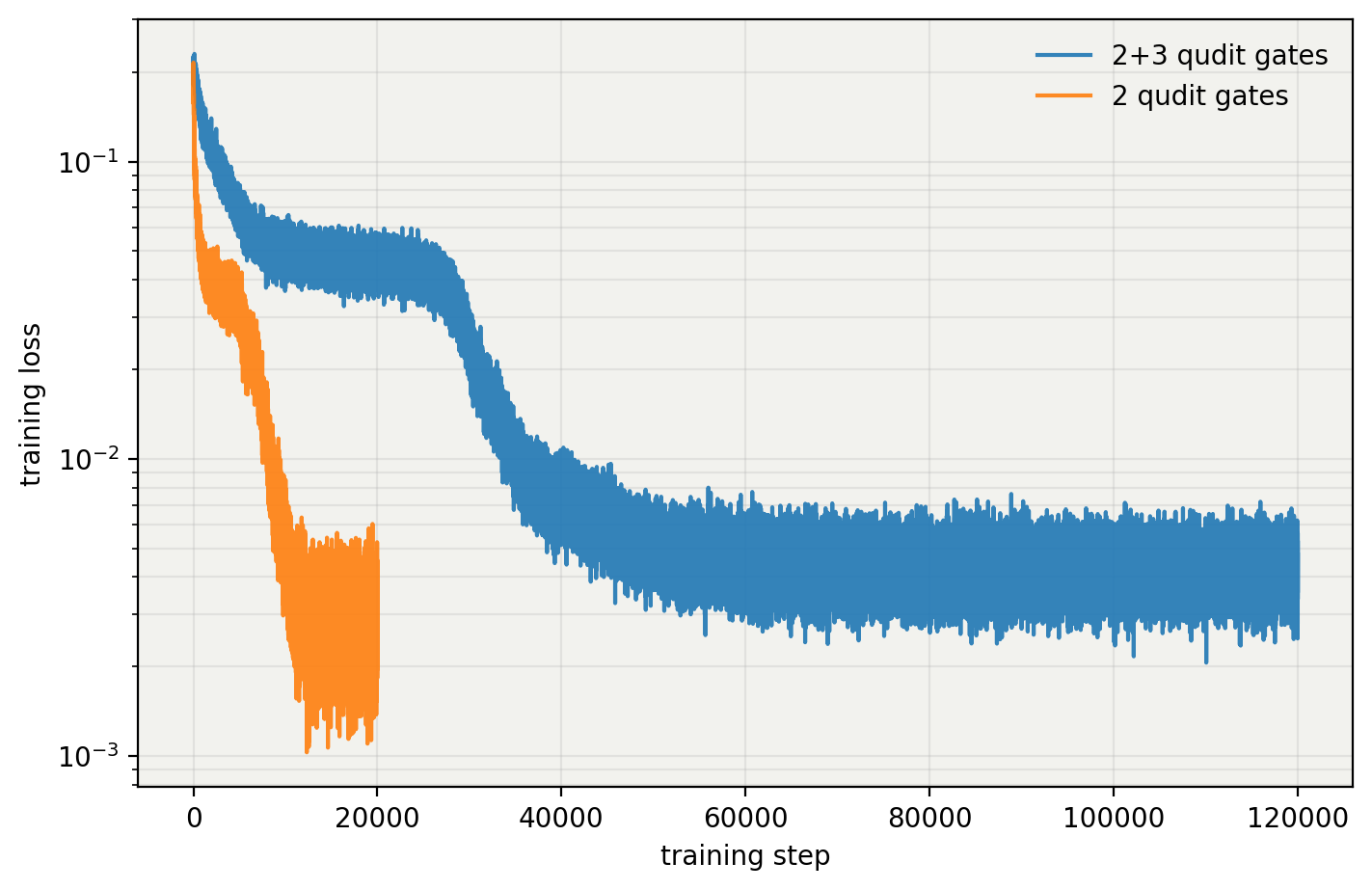}
    \caption{A spectral Born machine model with $\sim 40$K parameters (orange) converges faster on the rRNA dataset compared to a model $\sim 1.3M$ parameters (blue), though both models reach a similar loss. This suggests that inductive bias of the spectral Born machine permits good training performance across a large range of model complexities.}
    \label{fig:losses}
\end{figure}

Both models were trained with $|\mathcal{Z}|=|\mathcal{K}|=500$. Parameters were initialized in a similar way to the Potts model: single qudit gate parameters are chosen to prepare a uniform distribution, and all others are initialized with a normal distribution around zero with standard deviation $0.01$ (\textbf{model 1}) and $0.0001$ (\textbf{model 2}). Training model 1 took less than 3 minutes, whereas model 2 took close to 10 hours. In Fig.~\ref{fig:losses} we plot the training loss curves for the two runs. Both models train to similar values, with the smaller model converging to a slightly smaller loss in fewer iterations (0.0020 vs 0.0035). It is possible the larger model has not fully converged, however. We thus speculate that although model 2 has many more trainable parameters, they are not necessary since model 1 is already sufficiently expressive to capture the low order statistics. 

We visualize the trained model 2 by comparing its low order Fourier information with those of the combined training plus test empirical distribution (see Fig.~\ref{fig:rna_plots}). For each pair of non-trivial  $k_1,k_2\in\{1,2,3\}$ and sequence positions $i,j$, we plot the value of the Fourier coefficient $\k$, where the $i,j$ components of $\k$ are $(k_1,k_2)$ and zero otherwise. The diagonal of the plots show the order-1 Fourier coefficients, which are also shown explicitly below the 2D plots. 
Since Fourier coefficients are complex, colour hue records the phase of the coefficient while opacity records its magnitude. 
As suggested by the value of the loss, the close agreement between the empirical and model plots shows that the trained circuit has learned the dominant pairwise dependencies between sites in the Fourier basis, which are precisely the low-order spectral statistics targeted by the training loss. 

\section{Discussion}

\subsection{Spectral bias: expressivity in all the right places?}
It is well known that increased parameterization of neural network models can lead to better performance \cite{kaplan2020scaling}. However, this is often only possible when there is a sufficient abundance of data to prevent the model collapsing onto the training distribution. For example, in \cite{kaplan2020scaling}, the authors provide explicit guidance on \emph{``how much data we would need to train models of increasing size while keeping overfitting under control"}. 

This is one important area in which spectral Born machines may provide advantages relative to classical neural methods, since directly engineering a spectral bias towards low order Fourier concentration (as described in section \ref{sec:spectralbias}) can potentially act as a barrier to overfitting, and has even been suggested as key to understanding generalization in deep learning \cite{rahaman2019spectral, xu2019frequency}. The reason for this lies in uncertainty relations \cite{donoho1989uncertainty, meshulam2006uncertainty} between Fourier and direct (data) spaces: spectrally concentrated distributions are effectively sparse in Fourier space and thus should be dense in data space, whereas empirical training distributions are sparse in data space and therefore dense in Fourier space. If a model class has a strong bias towards low-order spectrally concentrated distributions, this can therefore forbid the possibility of memorization of training data, due to the inability to represent sparse distributions. Although further investigation is needed, the results of training on the rRNA dataset suggest this could be the case, since despite the model having close to four orders of magnitude more parameters than data, the training loss appears to suggest the model is not overfitted. 

We cautiously suggest that for many learning problems, spectral Born machines may therefore be parameterized in an advantageous way: extreme over-parameterization is possible to increase expressivity, but the increased expressivity is not the type that will overfit to the data. It is interesting to note that, in the classical literature, it has recently been suggested that powerful models could be precisely those that encode an order-dependent prior despite having the capacity to represent high order functions \cite{wilson2025deep}. Whether spectral Born machines are sufficiently flexible and correctly biased to fit this description is still unclear but is an exciting direction of future research. 

\subsection{Initialization is not a challenge}
Since gradients of spectral Born machines can only be estimated up to the statistical error determined by the number of samples $|\mathcal{Z}|$, they are equally susceptible to issues of barren plateaus as algorithms that train on quantum hardware. Despite this, as was the case with IQP Born machines \cite{recio2025train,shen2026characterizing, lerch2026iqp}, we did not encounter any problems related to model training. On the contrary, even when training the 1.3M parameter model, a simple zero-centered Gaussian initialization with relatively large standard deviation of $0.0001$ sufficed to quickly find a training signal and converge to a low loss value. This ease in finding suitable initializations goes counter to recent suggestions \cite{kulshrestha2026exponentially} that
\begin{quote}
    \itshape
    ``avoiding barren plateaus via smart initializations can trade the exponential concentration problem for the challenge of selecting the right trainable pocket amongst many options."
\end{quote}
Rather, given the scale at which our models are trained, we are inclined to suggest that the problem of model initialization is effectively solved, at least for this class of generative model. The reason for this, we speculate, is not particularly surprising or unique to spectral Born machines: initializing most parameters close to zero limits the spread of long-range correlations in the initial distribution, effectively encoding a prior belief of spectral concentration, and this starts training in a low complexity region of distribution space in which tractable gradients can be found. It is then not unreasonable to imagine that, as has been observed in neural networks \cite{chiang2023loss, liu2022loss}, the benefits of sufficient parameterization ensure a route to a local minimum that generalizes well, even if this happens to be one of many pockets.

\subsection{On dequantization}
Any promising quantum algorithm is at risk of potential dequantization, and train-on-classical approaches are now squarely in the cross-hairs of such attacks. A likely strategy for classically sampling from a spectral Born machine is to attempt to exploit the defining bias of the model, which is also a defining feature of many realistic distributions: information is concentrated primarily on low order Fourier coefficients. By exploiting a form of polynomial sparsity in the Fourier space, these techniques may thus aim to spoof the sampling step of the algorithm. 

While it is tempting to think this may be possible, there are also a number of reasons why one might expect this to fail. First, even though for a fixed Fourier weight cut-off, the number of Fourier coefficients is polynomial in the number of dits, the numbers quickly become very large. For example, for a $d=2$ model, considering all Fourier coefficients up to weight 10 requires keeping track of over $10^{23}$ numbers. Even when working with a tractable number of coefficients, one hits the issue that spectral Born machines only permit access to these numbers up to an inverse polynomial sampling error. Methods which aim to sample the distribution auto-regressively by sequentially computing conditional probabilities
\begin{align}
    p(x_j\vert x_{j-1}, x_{j-2},\cdots,x_1) = \frac{p(x_j, x_{j-1}, x_{j-2},\cdots,x_1)}{p(x_{j-1}, x_{j-2},\cdots,x_1)}
\end{align}
run into the problem that, although each $p$ on the right hand side has an effectively sparse representation in the Fourier basis, both of the probabilities are generally exponentially small, and inverse polynomial noise drowns out the ability to accurately estimate the ratio. 

Other approaches, such as constructing Gibbs states that sample the maximum entropy distribution given the spectral information, or using matrix product states to approximate the quantum state, are also not without challenges. In the former case due to the difficulty of finding the correct energy function to reproduce the low order marginals and difficulties in sampling from a thermal state, and in the latter case due to the exponential blow-up in bond dimension implied by densely connected phase gates. 

We believe it is very possible that spectral Born machines will inhabit a no man's land of computational complexity. On the one hand, efforts to classically dequantize the model could fail due to the above reasons, but on the other hand efforts to prove complexity theoretic statements in realistic parameter regimes could also fail, since household tools such as anticoncentration and random circuit sampling are not applicable. In our opinion, this is the space in which truly interesting models will live, and focusing excessive effort on classical simulability or complexity theoretic separations will limit our chances of finding truly groundbreaking approaches to machine learning. In the same way that classical machine learning researchers have grown used to the uncertainty surrounding why and how deep learning works just as well as it does, we may also need to grow used to the uncertainty of not knowing why or if a particular approach is classically hard in any precise sense. As always in machine learning, the ultimate proof of such advantages may have to come through impressive empirical results, which we may already be able to glimpse through the train-on-classical approach.

\section{Outlook}\label{sec:outlook}
There are a number of theoretical and practical challenges and opportunities that lie ahead. On the theoretical side, we still lack a proof of universality for spectral Born machines. In \cite{kurkin2025universality}, it was shown that any distribution over bitstrings of length $n$ can be realised as the marginal distribution of a IQP Born machine with $2n+1$ qubits. Since spectral Born machines contain IQP Born machines as a special case, by mapping dits to bits via a binary encoding it is possible to prepare any distribution over ditstrings in the same way as IQP Born machines. However, it would be interesting to understand if spectral Born machines are universal for distributions over ditstrings when explicitly working in the Fourier space of $\mathbb{Z}_d^n$, which may be possible by phasing the proof of \cite{kurkin2025universality} in more group Fourier theoretic language. More generally, such an understanding may also uncover theoretical insight into how the phase structure of the model connects to the space of parameterized distributions, which can further elucidate the bias of the model class. 

On a practical level, we appear to be hitting the limitations of what is possible with a single GPU. In order to develop truly scalable models that can reach the full scale of future quantum computers, it is therefore necessary to move to training on distributed GPU clusters. Luckily, this should be possible by making use of the same software and hardware advances that have enabled the deep learning revolution. Once this is possible, it is not unreasonable to expect training on billion parameter models to be within reach, and the true era of `quantum deep learning' can really begin. 

Finally, we emphasize that the methods and software we present can enable an entirely new approach to quantum machine learning that until recently has not been possible: large scale empirics. As we have seen with the numerical studies of this work, although sampling from the model is intractable with classical machines, the performance of trained models can to some extent still be probed by evaluation metrics that inspect the model in Fourier space. This opens the door to uncovering properties of quantum models that are unreachable with pen and paper analysis and could eventually find the ultimate reason why quantum computers can be a revolutionary technology for machine learning. 

\section{Acknowledgments}
We acknowledge Maria Schuld, Juan Miguel Arrazola and Alexia Salavrakos for discussion regarding the approach. AI tools were used in the project to aid with mathematical derivations and software implementations. Project notes, written in tandem with AI were then used to help create this manuscript, however the vast majority of the text was written by humans.

\bibliography{references.bib}

@inproceedings{rahaman2019spectral,
  title={On the spectral bias of neural networks},
  author={Rahaman, Nasim and Baratin, Aristide and Arpit, Devansh and Draxler, Felix and Lin, Min and Hamprecht, Fred and Bengio, Yoshua and Courville, Aaron},
  booktitle={{International Conference on Machine Learning}},
  pages={5301--5310},
  year={2019},
  organization={{PMLR}}
}

@article{xu2019frequency,
  title={Frequency principle: Fourier analysis sheds light on deep neural networks},
  author={Xu, Zhi-Qin John and Zhang, Yaoyu and Luo, Tao and Xiao, Yanyang and Ma, Zheng},
  journal={arXiv preprint arXiv:1901.06523},
  year={2019}
}

@inproceedings{
dherin2022why,
title={Why neural networks find simple solutions:  The many regularizers of geometric complexity},
author={Benoit Dherin and Michael Munn and Mihaela Rosca and David GT Barrett},
booktitle={Advances in Neural Information Processing Systems},
editor={Alice H. Oh and Alekh Agarwal and Danielle Belgrave and Kyunghyun Cho},
year={2022}}

@book{kondor2008group,
  title={Group theoretical methods in machine learning},
  author={Kondor, Imre Risi},
  year={2008},
  publisher={Columbia University}
}

@article{recio2025train,
  title={Train on classical, deploy on quantum: scaling generative quantum machine learning to a thousand qubits},
  author={Recio-Armengol, Erik and Ahmed, Shahnawaz and Bowles, Joseph},
  journal={arXiv preprint arXiv:2503.02934},
  year={2025}
}

@article{gretton2012kernel,
  title={A kernel two-sample test},
  author={Gretton, Arthur and Borgwardt, Karsten M and Rasch, Malte J and Sch{\"o}lkopf, Bernhard and Smola, Alexander},
  journal={The Journal of Machine Learning Research},
  volume={13},
  number={1},
  pages={723--773},
  year={2012},
  publisher={JMLR. org}
}

@article{liu2018differentiable,
  title={Differentiable learning of quantum circuit born machines},
  author={Liu, Jin-Guo and Wang, Lei},
  journal={Physical Review A},
  volume={98},
  number={6},
  pages={062324},
  year={2018},
  publisher={APS}
}

@article{rudolph2024trainability,
  title={Trainability barriers and opportunities in quantum generative modeling},
  author={Rudolph, Manuel S and Lerch, Sacha and Thanasilp, Supanut and Kiss, Oriel and Shaya, Oxana and Vallecorsa, Sofia and Grossi, Michele and Holmes, Zo{\"e}},
  journal={npj Quantum Information},
  volume={10},
  number={1},
  pages={116},
  year={2024},
  publisher={Nature Publishing Group UK London}
}

@article{wilson2025deep,
  title={Deep learning is not so mysterious or different},
  author={Wilson, Andrew Gordon},
  journal={arXiv preprint arXiv:2503.02113},
  year={2025}
}

@article{li2017mmd,
  title={Mmd gan: Towards deeper understanding of moment matching network},
  author={Li, Chun-Liang and Chang, Wei-Cheng and Cheng, Yu and Yang, Yiming and P{\'o}czos, Barnab{\'a}s},
  journal={Advances in Neural Information Processing Systems},
  volume={30},
  year={2017}
}

@article{hwobs,
  title={Heisenberg-Weyl Observables: Bloch vectors in phase space},
  author={Asadian, Ali and Erker, Paul and Huber, Marcus and Kl{\"o}ckl, Claude},
  journal={Physical Review A},
  volume={94},
  number={1},
  pages={010301},
  year={2016},
  publisher={APS}
}

@article{krebsbach2026encoding,
  title={Encoding Numerical Data for Generative Quantum Machine Learning},
  author={Krebsbach, Michael and Reiter, Florentin and Wellens, Thomas and Kowalski, Hagen-Henrik and Abedi, Ali},
  journal={arXiv preprint arXiv:2603.23407},
  year={2026}
}

@article{hoefler2023disentangling,
  title={Disentangling hype from practicality: On realistically achieving quantum advantage},
  author={Hoefler, Torsten and H{\"a}ner, Thomas and Troyer, Matthias},
  journal={Communications of the ACM},
  volume={66},
  number={5},
  pages={82--87},
  year={2023},
  publisher={ACM New York, NY, USA}
}

@article{bowles2025backpropagation,
  title={Backpropagation scaling in parameterised quantum circuits},
  author={Bowles, Joseph and Wierichs, David and Park, Chae-Yeun},
  journal={Quantum},
  volume={9},
  pages={1873},
  year={2025},
  publisher={Verein zur F{\"o}rderung des Open Access Publizierens in den Quantenwissenschaften}
}

@article{abbas2023quantum,
  title={On quantum backpropagation, information reuse, and cheating measurement collapse},
  author={Abbas, Amira and King, Robbie and Huang, Hsin-Yuan and Huggins, William J and Movassagh, Ramis and Gilboa, Dar and McClean, Jarrod},
  journal={Advances in Neural Information Processing Systems},
  volume={36},
  pages={44792--44819},
  year={2023}
}

@article{coyle2025training,
  title={Training-efficient density quantum machine learning},
  author={Coyle, Brian and Raj, Snehal and Mathur, Natansh and Cherrat, El Amine and Jain, Nishant and Kazdaghli, Skander and Kerenidis, Iordanis},
  journal={npj Quantum Information},
  volume={11},
  number={1},
  pages={172},
  year={2025},
  publisher={Nature Publishing Group UK London}
}

@article{chinzei2025trade,
  title={Trade-off between gradient measurement efficiency and expressivity in deep quantum neural networks},
  author={Chinzei, Koki and Yamano, Shinichiro and Tran, Quoc Hoan and Endo, Yasuhiro and Oshima, Hirotaka},
  journal={npj Quantum Information},
  volume={11},
  number={1},
  pages={79},
  year={2025},
  publisher={Nature Publishing Group UK London}
}

@article{mathur2026scalable,
  title={Scalable On-Hardware Training of Quantum Neural Networks and Application to Clinical Data Imputation},
  author={Mathur, Natansh and Barkoutsos, Panagiotis Kl and Yamada, Masako and Roetteler, Martin and Kerenidis, Iordanis},
  journal={arXiv preprint arXiv:2606.03517},
  year={2026}
}

@article{coyle2026adaptive,
  title={Adaptive directional gradients for parameterised quantum circuits},
  author={Coyle, Brian and Raj, Snehal and Umathe, Virag and Cherrat, El Amine and Kashefi, Elham},
  journal={arXiv preprint arXiv:2606.09734},
  year={2026}
}

@article{kolarovszki2026generative,
  title={Generative modeling with Gaussian Boson Sampling: classically trainable Bosonic Born Machines},
  author={Kolarovszki, Zolt{\'a}n and Bak{\'o}, Bence and Oszmaniec, Michal and Oh, Changhun and Zimbor{\'a}s, Zolt{\'a}n},
  journal={arXiv preprint arXiv:2603.11195},
  year={2026}
}

@article{bako2025fermionic,
  title={Fermionic Born Machines: Classical training of quantum generative models based on Fermion Sampling},
  author={Bak{\'o}, Bence and Kolarovszki, Zolt{\'a}n and Zimbor{\'a}s, Zolt{\'a}n},
  journal={arXiv preprint arXiv:2511.13844},
  year={2025}
}

@article{gottlieb2026efficient,
  title={Efficient training of photonic quantum generative models},
  author={Gottlieb, Felix and Mezher, Rawad and Ventura, Brian and Mansfield, Shane and Salavrakos, Alexia},
  journal={arXiv preprint arXiv:2603.08793},
  year={2026}
}

@article{belis2026spectral,
  title={Spectral methods: crucial for machine learning, natural for quantum computers?},
  author={Belis, Vasilis and Bowles, Joseph and Gupta, Rishabh and Peters, Evan and Schuld, Maria},
  journal={arXiv preprint arXiv:2603.24654},
  year={2026}
}

@article{dherin2022neural,
  title={Why neural networks find simple solutions: The many regularizers of geometric complexity},
  author={Dherin, Benoit and Munn, Michael and Rosca, Mihaela and Barrett, David},
  journal={Advances in Neural Information Processing Systems},
  volume={35},
  pages={2333--2349},
  year={2022}
}

@article{rosca2020case,
  title={A case for new neural network smoothness constraints},
  author={Rosca, Mihaela and Weber, Theophane and Gretton, Arthur and Mohamed, Shakir},
  journal= {PMLR},
  volume = {137},
  pages={21-32},
  year={2020},
}

@article{childs2010quantum,
  title={Quantum algorithms for algebraic problems},
  author={Childs, Andrew M and Van Dam, Wim},
  journal={Reviews of Modern Physics},
  volume={82},
  number={1},
  pages={1--52},
  year={2010},
  publisher={APS}
}

@article{armengol2025iqpopt,
  title={Iqpopt: Fast optimization of instantaneous quantum polynomial circuits in jax},
  author={Armengol, Erik and Bowles, Joseph},
  journal={arXiv preprint arXiv:2501.04776},
  year={2025}
}

@inproceedings{smola2003kernels,
  title={Kernels and regularization on graphs},
  author={Smola, Alexander J and Kondor, Risi},
  booktitle={Learning theory and kernel machines: 16th annual conference on learning theory and 7th kernel workshop, COLT/kernel 2003, Washington, DC, USA, august 24-27, 2003. Proceedings},
  pages={144--158},
  year={2003},
  organization={Springer}
}

@inproceedings{kondor2002diffusion,
  title={Diffusion kernels on graphs and other discrete structures},
  author={Kondor, Risi Imre and Lafferty, John},
  booktitle={Proceedings of the 19th international conference on machine learning},
  volume={2002},
  pages={315--322},
  year={2002}
}

@article{shuman2013emerging,
  title={The emerging field of signal processing on graphs: Extending high-dimensional data analysis to networks and other irregular domains},
  author={Shuman, David I and Narang, Sunil K and Frossard, Pascal and Ortega, Antonio and Vandergheynst, Pierre},
  journal={IEEE signal processing magazine},
  volume={30},
  number={3},
  pages={83--98},
  year={2013},
  publisher={IEEE}
}

@article{bergholm2018pennylane,
  title={Pennylane: Automatic differentiation of hybrid quantum-classical computations},
  author={Bergholm, Ville and Izaac, Josh and Schuld, Maria and Gogolin, Christian and Ahmed, Shahnawaz and Ajith, Vishnu and Alam, M Sohaib and Alonso-Linaje, Guillermo and AkashNarayanan, B and Asadi, Ali and others},
  journal={arXiv preprint arXiv:1811.04968},
  year={2018}
}

@article{kaplan2020scaling,
  title={Scaling laws for neural language models},
  author={Kaplan, Jared and McCandlish, Sam and Henighan, Tom and Brown, Tom B and Chess, Benjamin and Child, Rewon and Gray, Scott and Radford, Alec and Wu, Jeffrey and Amodei, Dario},
  journal={arXiv preprint arXiv:2001.08361},
  year={2020}
}

@article{wilson2020bayesian,
  title={Bayesian deep learning and a probabilistic perspective of generalization},
  author={Wilson, Andrew G and Izmailov, Pavel},
  journal={Advances in neural information processing systems},
  volume={33},
  pages={4697--4708},
  year={2020}
}

@article{lerch2026iqp,
  title={IQP born machines under data-dependent and agnostic initialization strategies},
  author={Lerch, Sacha and Bowles, Joseph and Puig, Ricard and Armengol, Erik and Holmes, Zo{\"e} and Thanasilp, Supanut},
  journal={arXiv preprint arXiv:2603.14576},
  year={2026}
}

@article{shen2026characterizing,
  title={Characterizing trainability of instantaneous quantum polynomial circuit born machines},
  author={Shen, Kevin and Pielawa, Susanne and Dunjko, Vedran and Wang, Hao},
  journal={arXiv preprint arXiv:2602.11042},
  year={2026}
}

@article{kurkin2025universality,
  title={Universality and kernel-adaptive training for classically trained, quantum-deployed generative models},
  author={Kurkin, Andrii and Shen, Kevin and Pielawa, Susanne and Wang, Hao and Dunjko, Vedran},
  journal={arXiv preprint arXiv:2510.08476},
  year={2025}
}

@article{kurkin2026universality,
  title={Universality of Classically Trainable, Quantum-Deployed Boson-Sampling Generative Models},
  author={Kurkin, Andrii and Chabaud, Ulysse and Kolarovszki, Zolt{\'a}n and Bak{\'o}, Bence and Zimbor{\'a}s, Zolt{\'a}n and Dunjko, Vedran},
  journal={arXiv preprint arXiv:2603.11014},
  year={2026}
}

@article{kulshrestha2026exponentially,
  title={Exponentially many initializations to avoid barren plateaus},
  author={Kulshrestha, Ankit and Puig, Ricard and Garc{\'\i}a-Mart{\'\i}n, Diego and Cincio, Lukasz and Safro, Ilya and Holmes, Zo{\"e} and Cerezo, M},
  journal={arXiv preprint arXiv:2606.18515},
  year={2026}
}

@misc{tcdq_note,
  title  = {},
  author = {},
  year   = {},
  note   = {The tcdq module is currently under development but available as part of PennyLane labs. See \url{https://docs.pennylane.ai/en/latest/code/qp_labs.html}.}
}

@software{jax2018github,
  author = {James Bradbury and Roy Frostig and Peter Hawkins and Matthew James Johnson and Yash Katariya and Chris Leary and Dougal Maclaurin and George Necula and Adam Paszke and Jake Vander{P}las and Skye Wanderman-{M}ilne and Qiao Zhang},
  title = {{JAX}: composable transformations of {P}ython+{N}um{P}y programs},
  version = {0.3.13},
  year = {2018},
}

@article{banks2026qudit,
  title={Qudit extension of parameterized IQP circuits: A generative quantum machine learning approach to integer data},
  author={Banks, Robert J and Crippa, Arianna and Traube, Matthias and Unger, Josua and Ertler, Christian and Lechner, Wolfgang},
  journal={arXiv preprint arXiv:2606.28236},
  year={2026}
}

@article{wu1982potts,
  title={The potts model},
  author={Wu, Fa-Yueh},
  journal={Reviews of modern physics},
  volume={54},
  number={1},
  pages={235},
  year={1982},
  publisher={APS}
}

@article{griffiths2003rfam,
  title={Rfam: an RNA family database},
  author={Griffiths-Jones, Sam and Bateman, Alex and Marshall, Mhairi and Khanna, Ajay and Eddy, Sean R},
  journal={Nucleic acids research},
  volume={31},
  number={1},
  pages={439--441},
  year={2003},
  publisher={Oxford University Press}
}

@article{herrero2025born,
  title={The Born Ultimatum: Conditions for Classical Surrogation of Quantum Generative Models with Correlators},
  author={Herrero-Gonzalez, Mario and Coyle, Brian and McDowall, Kieran and Grassie, Ross and Beentjes, Sjoerd and Khamseh, Ava and Kashefi, Elham},
  journal={arXiv preprint arXiv:2511.01845},
  year={2025}
}

@article{ballo2026shallow,
  title={Shallow instantaneous quantum polynomial-time circuits for generative modeling on noisy intermediate-scale quantum hardware},
  author={Ball{\'o}-Gimbernat, Oriol and Arroyo-S{\'a}nchez, Marcos and Garc{\'\i}a-Molina, Paula and Garriga, Adan and Vilari{\~n}o, Fernando},
  journal={Physical Review A},
  volume={113},
  number={4},
  pages={042617},
  year={2026},
  publisher={APS}
}

@article{herbst2025limits,
  title={Limits of quantum generative models with classical sampling hardness},
  author={Herbst, Sabrina and Brandi{\'c}, Ivona and P{\'e}rez-Salinas, Adri{\'a}n},
  journal={arXiv preprint arXiv:2512.24801},
  year={2025}
}

@article{slim2026iqp,
  title={An IQP Born Machine for Calorimeter Image Generation at 64 Qubits with Compiled-IQP Deployment},
  author={Slim, Jamal and Monaco, Saverio and Rehm, Florian and Kr{\"u}cker, Dirk and Borras, Kerstin},
  journal={arXiv preprint arXiv:2605.27735},
  year={2026}
}

@article{baumann2026parity,
  title={Parity Supervision as a Driver of Generalization in Quantum Generative Modeling},
  author={Baumann, Markus and Hein, Daniel and Udluft, Steffen and Rohe, Tobias and Linnhoff-Popien, Claudia and Stein, Jonas},
  journal={arXiv preprint arXiv:2605.10258},
  year={2026}
}

@article{marshall2024improved,
  title={Improved separation between quantum and classical computers for sampling and functional tasks},
  author={Marshall, Simon C and Aaronson, Scott and Dunjko, Vedran},
  journal={arXiv preprint arXiv:2410.20935},
  year={2024}
}

@article{bremner2011classical,
  title={Classical simulation of commuting quantum computations implies collapse of the polynomial hierarchy},
  author={Bremner, Michael J and Jozsa, Richard and Shepherd, Dan J},
  journal={Proceedings of the Royal Society A: Mathematical, Physical and Engineering Sciences},
  volume={467},
  number={2126},
  pages={459--472},
  year={2011},
  publisher={The Royal Society Publishing}
}

@article{bremner2016average,
  title={Average-case complexity versus approximate simulation of commuting quantum computations},
  author={Bremner, Michael J and Montanaro, Ashley and Shepherd, Dan J},
  journal={Physical review letters},
  volume={117},
  number={8},
  pages={080501},
  year={2016},
  publisher={APS}
}

@article{meshulam2006uncertainty,
  title={An uncertainty inequality for finite abelian groups},
  author={Meshulam, Roy},
  journal={European Journal of Combinatorics},
  volume={27},
  number={1},
  pages={63--67},
  year={2006},
  publisher={Elsevier}
}

@article{donoho1989uncertainty,
  title={Uncertainty principles and signal recovery},
  author={Donoho, David L. and Stark, Philip B.},
  journal={SIAM Journal on Applied Mathematics},
  volume={49},
  number={3},
  pages={906--931},
  year={1989},
  publisher={SIAM}}

@article{xu2018empirical,
  title={An empirical study on evaluation metrics of generative adversarial networks},
  author={Xu, Qiantong and Huang, Gao and Yuan, Yang and Guo, Chuan and Sun, Yu and Wu, Felix and Weinberger, Kilian},
  journal={arXiv preprint arXiv:1806.07755},
  year={2018}
}

@article{theis2015note,
  title={A note on the evaluation of generative models},
  author={Theis, Lucas and Oord, A{\"a}ron van den and Bethge, Matthias},
  journal={arXiv preprint arXiv:1511.01844},
  year={2015}
}

@article{bischoff2024practical,
  title={A practical guide to sample-based statistical distances for evaluating generative models in science},
  author={Bischoff, Sebastian and Darcher, Alana and Deistler, Michael and Gao, Richard and Gerken, Franziska and Gloeckler, Manuel and Haxel, Lisa and Kapoor, Jaivardhan and Lappalainen, Janne K and Macke, Jakob H and others},
  journal={arXiv preprint arXiv:2403.12636},
  year={2024}
}

@inproceedings{chiang2023loss,
  title={Loss landscapes are all you need: Neural network generalization can be explained without the implicit bias of gradient descent},
  author={Chiang, Ping-yeh and Ni, Renkun and Miller, David Yu and Bansal, Arpit and Geiping, Jonas and Goldblum, Micah and Goldstein, Tom},
  booktitle={The Eleventh International Conference on Learning Representations},
  year={2023}
}

@article{liu2022loss,
  title={Loss landscapes and optimization in over-parameterized non-linear systems and neural networks},
  author={Liu, Chaoyue and Zhu, Libin and Belkin, Mikhail},
  journal={Applied and Computational Harmonic Analysis},
  volume={59},
  pages={85--116},
  year={2022},
  publisher={Elsevier}
}

@article{kasture2023protocols,
  title={Protocols for classically training quantum generative models on probability distributions},
  author={Kasture, Sachin and Kyriienko, Oleksandr and Elfving, Vincent E},
  journal={Physical Review A},
  volume={108},
  number={4},
  pages={042406},
  year={2023},
  publisher={APS}
}

@article{tuysuz2026quantum,
  title={Quantum Fourier Generative Models Trainable at Large Scale},
  author={T{\"u}ys{\"u}z, Cenk and Kyriienko, Oleksandr and Grossi, Michele},
  journal={arXiv preprint arXiv:2606.28483},
  year={2026}
}

\appendix

\section{Proof of Proposition~\ref{prop:zd_classical}}\label{ap:hw-estimate}

Using the form $U(\btheta) = (F^{\otimes n})^{\dagger}D(\btheta)F^{\otimes n}$ from Sec.~\ref{sec:circuit_class} together with the conjugation relation \eqref{eq:F_conjugation}, which implies $F^{\otimes n}\,\mathcal{D}(\k,\m)\,(F^{\otimes n})^{\dagger} = \mathcal{D}(\m,-\k)$, we have
\begin{align}\label{eq:hw_proof_step1}
    \langle\mathcal{D}(\k,\m)\rangle_{\btheta} = \bra{+}D^{\dagger}(\btheta)\,\mathcal{D}(\m,-\k)\,D(\btheta)\ket{+},
\end{align}
where $\ket{+} = d^{-n/2}\sum_{\z\in\mathbb{Z}_d^n}\ket{\z}$ is the equal superposition over $\mathbb{Z}_d^n$. From the definitions \eqref{eq:clockphase} of the shift and clock operators, the action of the displacement operator on a computational-basis state is
\begin{align}\label{eq:hw_action}
    \mathcal{D}(\m,-\k)\ket{\z} = \exp\!\left(\frac{i\pi}{d}\,\m\cdot(2\z-\k)\right)\ket{\z\ominus\k},
\end{align}
so that for any $\z'\in\mathbb{Z}_d^n$, $\bra{\z'}\mathcal{D}(\m,-\k)\ket{\z} = \delta_{\z',\z\ominus\k}\,\exp\!\left(\frac{i\pi}{d}\,\m\cdot(2\z-\k)\right)$. Inserting two resolutions of the identity in the computational basis into \eqref{eq:hw_proof_step1} and using \eqref{eq:Dunitary}, the Kronecker delta in the matrix element collapses one of the sums and we obtain
\begin{multline}
    \langle\mathcal{D}(\k,\m)\rangle_{\btheta} = \\ \frac{1}{d^n}\sum_{\z\in\mathbb{Z}_d^n}\exp\!\left(\frac{i\pi}{d}\,\m\cdot(2\z-\k)\right)\exp\!\big(i\big[\Phi_{\btheta}(\z) - \Phi_{\btheta}(\z\ominus\k)\big]\big).
\end{multline}
Since $1/d^n$ is the uniform probability over $\mathbb{Z}_d^n$ and using the definition \eqref{eq:phase_diff} of the accumulated phase difference $\Delta^{\k}_{\btheta}(\z)$, we find \eqref{eq:hw_estimator}:
\begin{align}\nonumber
    \langle\mathcal{D}(\k,\m)\rangle_{\btheta} = \mathbb{E}_{\z\sim \text{Unif}}\!\left[\exp\!\left(\frac{i\pi}{d}\,\m\cdot(2\z-\k)+i\,\Delta^{\k}_{\btheta}(\z)\right)\right],
\end{align}
Since the integrand is a unit-modulus complex number, the empirical mean
\begin{align}\label{eq:hw_sample_mean}
    \widehat{\langle\mathcal{D}(\k,\m)\rangle}_{\btheta} = \frac{1}{|\mathcal{Z}|}\sum_{\z\in\mathcal{Z}}\exp\!\left(\frac{i\pi}{d}\,\m\cdot(2\z-\k)+i\,\Delta^{\k}_{\btheta}(\z)\right)
\end{align}
formed from samples $\mathcal{Z} = \{\z_1,\ldots,\z_{|\mathcal{Z}|}\}\sim \text{Unif}$ is an unbiased estimator of $\langle\mathcal{D}(\k,\m)\rangle_{\btheta}$ whose real and imaginary parts each have standard deviation bounded by $1/\sqrt{|\mathcal{Z}|}$. Taking $|\mathcal{Z}|\ge 1/\epsilon^2$ therefore yields an estimator with error at most $\epsilon$, requiring $\mathcal{O}(1/\epsilon^2)$ samples. By the assumption of Proposition~\ref{prop:zd_classical} that $\Phi_{\btheta}$ is computable in $\mathrm{poly}(n,d)$ time and space, each evaluation of $\Delta^{\k}_{\btheta}(\z)$ reduces to two such evaluations of $\Phi_{\btheta}$, giving a total runtime of $\mathrm{poly}(n,d)/\epsilon^2$ in $\mathrm{poly}(n,d)$ space.

Because the summands have unit modulus, their real and imaginary parts lie in $[-1,1]$, and we can strengthen this to a high-probability guarantee. Applying Hoeffding's inequality to each part with deviation $\epsilon/\sqrt{2}$ and taking a union bound, the total error $\big|\widehat{\langle\mathcal{D}(\k,\m)\rangle}_{\btheta} - \langle\mathcal{D}(\k,\m)\rangle_{\btheta}\big|$ exceeds $\epsilon$ with probability at most $4\exp(-|\mathcal{Z}|\epsilon^2/4)$. Hence $|\mathcal{Z}| \ge (4/\epsilon^2)\ln(4/\delta)$ samples suffice for additive error $\epsilon$ with probability at least $1-\delta$.

\section{Derivation of the batched matrix-multiplication estimator}\label{ap:batching}

In this appendix we derive the matrix expressions \eqref{eq:hw_batch}, \eqref{eq:Jmatrix}, and \eqref{eq:factored_E}. We are given the batch of expectation values
\begin{align}
    \mathcal{O}_i = \langle\mathcal{D}(\k_i,\m_i)\rangle_{\btheta}, \qquad i=1,\ldots,|\mathcal{O}|,
\end{align}
to be estimated using a single batch of Monte-Carlo samples $\mathcal{Z}=\{\z_i\}$ drawn uniformly from $\mathbb{Z}_d^n$. Stacking the parameters of the observables and the samples row-wise as the integer matrices
\begin{align}
    K &= \begin{pmatrix}\k_1\\ \vdots\\ \k_{|\mathcal{O}|}\end{pmatrix} \in \mathbb{Z}_d^{|\mathcal{O}|\times n}, \\[5pt] M &= \begin{pmatrix}\m_1\\ \vdots\\ \m_{|\mathcal{O}|}\end{pmatrix} \in \mathbb{Z}_d^{|\mathcal{O}|\times n}, \\[5pt] Z &= \begin{pmatrix}\z_1\\ \vdots\\ \z_{|\mathcal{Z}|}\end{pmatrix} \in \mathbb{Z}_d^{|\mathcal{Z}|\times n},
\end{align}
the empirical estimator \eqref{eq:hw_sample_mean} of $\mathcal{O}_i$ reads
\begin{align}\label{eq:hw_per_sample}
    \widehat{\mathcal{O}}_i = \frac{1}{|\mathcal{Z}|}\sum_{\z\in\mathcal{Z}}\exp\!\big(i\,J_{i,\z} + i\,E_{i,\z}\big),
\end{align}
with the prefactor and accumulated-phase entries
\begin{align}
    J_{i,\z} &= \frac{\pi}{d}\,\m_i\cdot(2\z - \k_i), \\
    E_{i,\z} &= \Delta^{\k_i}_{\btheta}(\z) = \Phi_{\btheta}(\z) - \Phi_{\btheta}(\z\ominus\k_i),
\end{align}
forming $|\mathcal{O}|\times|\mathcal{Z}|$ matrices $J$ and $E$. Stacking the empirical estimators into a column vector and applying the element-wise exponential, the entire batch is computed as
\begin{align}
    (\widehat{\mathcal{O}}_1,\ldots,\widehat{\mathcal{O}}_{|\mathcal{O}|})^T = \frac{1}{|\mathcal{Z}|}\,\exp\!\big(i(J + E)\big)\,\boldsymbol{1}_{|\mathcal{Z}|},
\end{align}
which gives \eqref{eq:hw_batch}. Below we derive the matrix expressions for $J$ and $E$ that allow them to be assembled efficiently.

\subsection{The prefactor matrix \texorpdfstring{$J$}{J}}
Expanding the entry $J_{i,\z}$,
\begin{align}
    J_{i,\z} = \frac{\pi}{d}\big[2\,\m_i\cdot\z - \m_i\cdot\k_i\big].
\end{align}
The bilinear term $2\,\m_i\cdot\z$ is the $(i,\z)$ entry of $2MZ^T \in \mathbb{Z}^{|\mathcal{O}|\times|\mathcal{Z}|}$. The term $\m_i\cdot\k_i$ depends only on $i$ and is the $i$-th entry of $(M\odot K)\boldsymbol{1}_n$, which we broadcast across $|\mathcal{Z}|$ columns by right-multiplying with $\boldsymbol{1}_{|\mathcal{Z}|}^T$. This recovers \eqref{eq:Jmatrix}:
\begin{align}
    J = \frac{\pi}{d}\big(2 M Z^T - ((M\odot K)\boldsymbol{1}_n)\,\boldsymbol{1}_{|\mathcal{Z}|}^T\big).
\end{align}

\subsection{The phase matrix \texorpdfstring{$E$}{E}}
The matrix $E$ is more involved. Using the explicit form \eqref{eq:phi_HW} of the phase function $\Phi_{\btheta}(\z) = \sum_{\g\in\mathcal{G}}\theta_{\g}\,\phi_{\g}(\z)$, the entries decompose as
\begin{align}
    E_{i,\z} = \sum_{\g\in\mathcal{G}}\theta_{\g}\big(\phi_{\g}(\z) - \phi_{\g}(\z\ominus\k_i)\big).
\end{align}
Splitting the gate set $\mathcal{G}$ into subsets $\mathcal{G}_w$ of generators with weight $w$ (i.e.\ exactly $w$ non-zero entries), $\mathcal{G} = \bigsqcup_{w}\mathcal{G}_w$, the matrix decomposes additively as $E = \sum_w E_w$, where
\begin{align}
    [E_w]_{i,\z} = \sum_{\g\in\mathcal{G}_w}\theta_{\g}\big(\phi_{\g}(\z) - \phi_{\g}(\z\ominus\k_i)\big).
\end{align}
We will show that each $E_w$ admits an expansion as a sum of $2^w + 1$ triple matrix products.

Recalling the per-gate eigenvalue \eqref{eq:phigz},
\begin{align}
    \phi_{\g}(\z) = \prod_{j=1}^n c(g_j,z_j), \qquad c(v,u) := \sqrt{2}\cos\!\left(\frac{2\pi v u}{d}+\frac{\pi}{4}\right),
\end{align}
and using $c(0,u) = 1$, the eigenvalue is supported on $S_{\g} := \{j : g_j \ne 0\}$ and reads $\phi_{\g}(\z) = \prod_{j\in S_{\g}} c(g_j,z_j)$. Define the auxiliary function
\begin{align}
    \bar{s}(v,u) := \sqrt{2}\sin\!\left(\frac{2\pi v u}{d}+\frac{\pi}{4}\right).
\end{align}
The angle-addition formula then gives \small
\begin{align}\label{eq:angle_add_app}
    c(v,u-k) = c(v,u)\cos\!\left(\frac{2\pi v k}{d}\right) + \bar{s}(v,u)\sin\!\left(\frac{2\pi v k}{d}\right).
\end{align}\normalsize
Defining
\begin{align}
    T_0(v,u) &= c(v,u), & T_1(v,u) &= \bar{s}(v,u), \\
    A_0(v,k) &= \cos\!\left(\tfrac{2\pi v k}{d}\right), & A_1(v,k) &= -\sin\!\left(\tfrac{2\pi v k}{d}\right),
\end{align}
the identity \eqref{eq:angle_add_app} reads $c(v,u-k) = T_0(v,u)A_0(v,k) - T_1(v,u)A_1(v,k)$. Applying this to each factor of $\phi_{\g}(\z\ominus\k)$ and expanding the product over $\boldsymbol{\sigma}\in\{0,1\}^w$ yields
\begin{align}\nonumber
    \phi_{\g}(\z\ominus\k) &= \prod_{r=1}^w\!\big[T_0\,A_0 - T_1\,A_1\big](g_{S_{\g}(r)},z_{S_{\g}(r)},k_{S_{\g}(r)}) \\ 
    &= \sum_{\boldsymbol{\sigma}\in\{0,1\}^w}(-1)^{|\boldsymbol{\sigma}|}\prod_{r=1}^w T_{\sigma_r}\,A_{\sigma_r},
\end{align}
where the arguments of each factor are $(g_{S_{\g}(r)},z_{S_{\g}(r)})$ for $T_{\sigma_r}$ and $(g_{S_{\g}(r)},k_{S_{\g}(r)})$ for $A_{\sigma_r}$. Each term in the expansion factorises into a product of functions of $(\g,\z)$ times a product of functions of $(\g,\k)$, which is the structure required for a matrix formulation.

Concretely, define the matrices
\begin{align}
    [\mathbf{B}^w_{\boldsymbol{\sigma}}]_{\g,\z} &= \prod_{r=1}^w T_{\sigma_r}(g_{S_{\g}(r)},z_{S_{\g}(r)}), \\
    [\mathbf{C}^w_{\boldsymbol{\sigma}}]_{\g,\k} &= \prod_{r=1}^w A_{\sigma_r}(g_{S_{\g}(r)},k_{S_{\g}(r)}),
\end{align}
of shapes $|\mathcal{G}_w|\times|\mathcal{Z}|$ and $|\mathcal{G}_w|\times|\mathcal{O}|$ respectively, together with the $|\mathcal{G}_w|\times|\mathcal{Z}|$ matrix
\begin{align}
    [\widetilde{\mathbf{B}}_w]_{\g,\z} = \prod_{r=1}^w c(g_{S_{\g}(r)},z_{S_{\g}(r)}),
\end{align}
which encodes the unshifted eigenvalue $\phi_{\g}(\z)$. Writing $\boldsymbol{\theta}_w$ for the column vector of parameters $\{\theta_{\g}\}_{\g\in\mathcal{G}_w}$, the contribution
\begin{align}
    [E_w]_{i,\z} = \sum_{\g\in\mathcal{G}_w}\theta_{\g}\,\phi_{\g}(\z) - \sum_{\g\in\mathcal{G}_w}\theta_{\g}\,\phi_{\g}(\z\ominus\k_i)
\end{align}
becomes, in matrix form,
\begin{align}\nonumber
    E_w = \boldsymbol{1}_{|\mathcal{O}|}\boldsymbol{\theta}_w^T\widetilde{\mathbf{B}}_w \;-\; \sum_{\boldsymbol{\sigma}\in\{0,1\}^w}(-1)^{|\boldsymbol{\sigma}|}\,(\mathbf{C}^w_{\boldsymbol{\sigma}})^T\,\mathrm{diag}(\boldsymbol{\theta}_w)\,\mathbf{B}^w_{\boldsymbol{\sigma}},
\end{align}
which after relabelling $(-1)^{|\boldsymbol{\sigma}|+1} = -(-1)^{|\boldsymbol{\sigma}|}$ give \eqref{eq:factored_E}. The full phase matrix is then $E = \sum_w E_w$.

\subsection{Total complexity}
For a gate set restricted to weight-$w$ generators, building $E_w$ requires assembling each of the $2^w$ pairs of matrices $(\mathbf{B}^w_{\boldsymbol{\sigma}},\mathbf{C}^w_{\boldsymbol{\sigma}})$ at cost $\mathcal{O}(|\mathcal{G}_w|(|\mathcal{Z}|+|\mathcal{O}|))$ each, plus the matrix $\widetilde{\mathbf{B}}_w$ at cost $\mathcal{O}(|\mathcal{G}_w||\mathcal{Z}|)$. Performing the corresponding triple matrix multiplications then costs $\mathcal{O}(|\mathcal{G}_w||\mathcal{O}||\mathcal{Z}|)$ per term. Adding the cost $\mathcal{O}(|\mathcal{O}||\mathcal{Z}|n)$ of building $J$ from the $MZ^T$ multiplication, the total runtime to produce the batch \eqref{eq:hw_batch} is
\begin{align}\label{eq:total_complexity}
    O\!\,\Big(|\mathcal{O}|\,|\mathcal{Z}|\,n \;+\; \sum_{w} 2^w\,|\mathcal{G}_w|\big((|\mathcal{O}|+|\mathcal{Z}|) + |\mathcal{O}||\mathcal{Z}|\big)\Big).
\end{align}
The runtime is linear in each of $|\mathcal{O}|$, $|\mathcal{Z}|$, $|\mathcal{G}_w|$ and $n$ when the others are held fixed.

\section{Estimating the MMD loss from finite samples}\label{ap:mmd-estimation}

In this appendix we describe how to estimate the MMD loss \eqref{eq:zd_mmd_mixture} from finite samples by deriving an unbiased U-statistic estimator and show that it admits an efficient matrix-form implementation. 

\begin{proposition}[Unbiased MMD estimator]\label{prop:mmd_unbiased}
Let $\mathcal{X}=\{\x_i\sim p\}$, $\mathcal{K}=\{\k_j\sim\mathcal{P}_t\}$, and $\mathcal{Z}=\{\z_r\sim \text{Unif}\}$ be independent batches. Then the U-statistic estimator $\widehat{\mathrm{MMD}^2_{\mathrm{u}}}(\mathcal{X},\mathcal{K}, \mathcal{Z},\btheta)$ defined in \eqref{eq:mmd_unbiased} is unbiased:
\begin{align}
    \mathbb{E}_{\mathcal{X},\mathcal{K},\mathcal{Z}}\!\left[\widehat{\mathrm{MMD}^2_{\mathrm{u}}}(\mathcal{X},\mathcal{K}, \mathcal{Z},\btheta)\right] = \mathrm{MMD}^2(p,q_{\btheta}).
\end{align}
\end{proposition}

The bias is removed by a U-statistic construction analogous to the one used for parameterised IQP circuits in \cite{recio2025train}. Recalling from \eqref{eq:hw_estimator} (specialized to $\m = \boldsymbol{0}$) that the model expectation admits the classical Monte-Carlo form \small
\begin{align}
    \langle\mathcal{D}(\k,0)\rangle_{q_{\btheta}} &= \mathbb{E}_{\z\sim \text{Unif}}\!\big[f_{q_{\btheta}}(\k,\z,\btheta)\big], \\ f_{q_{\btheta}}(\k,\z,\btheta) :&= \exp\!\big(i\,\Delta^{\k}_{\btheta}(\z)\big). 
\end{align}\normalsize
From data samples we similarly have
\begin{align}\label{eq:fp_zd}
    \langle\mathcal{D}(\k,0)\rangle_{p} = \mathbb{E}_{\x\sim p}\!\big[f_p(\k,\x)\big], \quad f_p(\k,\x) := \omega^{\k\cdot\x}.
\end{align}

Writing $\mu_r(\k) := \langle\mathcal{D}(\k,0)\rangle_r$ for $r\in\{p,q_{\btheta}\}$ and expanding the squared modulus in \eqref{eq:zd_mmd_mixture} gives three terms,
\begin{multline}\label{eq:three_terms}
    \big|\mu_p(\k) - \mu_{q_{\btheta}}(\k)\big|^2 = \mu_p^*(\k)\mu_p(\k) \\ - 2\,\mathrm{Re}\!\big[\mu_p^*(\k)\mu_{q_{\btheta}}(\k)\big] + \mu_{q_{\btheta}}^*(\k)\mu_{q_{\btheta}}(\k).
\end{multline}
Each of $\mu_p$ and $\mu_{q_{\btheta}}$ is itself a single expectation, so the squared terms are products of two independent expectations and require the standard ``remove the diagonal'' trick to be unbiasedly estimated from finite samples. The cross term $\mu_p^*\mu_{q_{\btheta}}$ involves \emph{independent} batches of $\x$ and $\z$ samples and so requires no diagonal removal. Concretely, given batches $\mathcal{X} = \{\x_i\sim p\}$, $\mathcal{K} = \{\k_j\sim\mathcal{P}\}$, $\mathcal{Z} = \{\z_r\sim \text{Unif}\}$, unbiased estimators of $\mathbb{E}_{\k\sim\mathcal{P}}[\,\cdot\,]$ for each of the three terms in \eqref{eq:three_terms} are
\begin{align}
    \widehat{\mathbb{E}_{\k}[\mu_p^*\mu_p]} &= \frac{1}{|\mathcal{K}|\,|\mathcal{X}|(|\mathcal{X}|-1)}\sum_{j}\sum_{i\ne i'}\omega^{\k_j\cdot(\x_{i'}-\x_i)},\label{eq:zd_term_pp}\\[2pt]
    \widehat{\mathbb{E}_{\k}[\mu_p^*\mu_{q_{\btheta}}]} &= \frac{1}{|\mathcal{K}|\,|\mathcal{X}|\,|\mathcal{Z}|}\sum_{i,j,r}\omega^{-\k_j\cdot\x_i}\,f_{q_{\btheta}}(\k_j,\z_r,\btheta),\label{eq:zd_term_pq}
\end{align}
and
\begin{multline}
    \widehat{\mathbb{E}_{\k}[\mu_{q_{\btheta}}^*\mu_{q_{\btheta}}]}   = \\  \frac{1}{|\mathcal{K}|\,|\mathcal{Z}|(|\mathcal{Z}|-1)}\sum_{j}\sum_{r\ne r'}f^*_{q_{\btheta}}(\k_j,\z_r,\btheta)\,f_{q_{\btheta}}(\k_j,\z_{r'},\btheta).\label{eq:zd_term_qq}
\end{multline}
Note that \eqref{eq:zd_term_pp} and \eqref{eq:zd_term_qq} are automatically real (each is of the form $|\sum_i a_i|^2 - \sum_i|a_i|^2$ before normalisation); only the cross term \eqref{eq:zd_term_pq} is in general complex, and we take its real part. Combining these gives the $\mathbb{Z}_d^n$ unbiased estimator
\begin{multline}\label{eq:mmd_unbiased}
    \widehat{\mathrm{MMD}^2_{\mathrm{u}}}(\mathcal{X},\mathcal{K},\mathcal{Z},\btheta) = \frac{1}{|\mathcal{K}|\,|\mathcal{X}|(|\mathcal{X}|-1)}\sum_{j}\sum_{i\ne i'}\omega^{\k_j\cdot(\x_{i'}-\x_i)}\\
    -\frac{2}{|\mathcal{K}|\,|\mathcal{X}|\,|\mathcal{Z}|}\,\mathrm{Re}\!\sum_{i,j,r}\omega^{-\k_j\cdot\x_i}\,f_{q_{\btheta}}(\k_j,\z_r,\btheta)\\
    +\frac{1}{|\mathcal{K}|\,|\mathcal{Z}|(|\mathcal{Z}|-1)}\sum_{j}\sum_{r\ne r'}f^*_{q_{\btheta}}(\k_j,\z_r,\btheta)\,f_{q_{\btheta}}(\k_j,\z_{r'},\btheta),
\end{multline}
satisfying $\mathbb{E}_{\mathcal{X},\mathcal{K},\mathcal{Z}}[\widehat{\mathrm{MMD}^2_{\mathrm{u}}}] = \mathrm{MMD}^2(p,q_{\btheta})$. As in \cite{recio2025train}, for any fixed $\k_j$, the constraints $i\ne i'$ and $r\ne r'$ ensure that the paired draws are i.i.d.\ when averaged over the sampling of $\mathcal{X}$ and $\mathcal{Z}$ respectively, so each pairwise sum estimates the corresponding product of expectations. 

To estimate the loss efficiently, we stack the batches as integer arrays $X\in\mathbb{Z}_d^{|\mathcal{X}|\times n}$, $K\in\mathbb{Z}_d^{|\mathcal{K}|\times n}$, $Z\in\mathbb{Z}_d^{|\mathcal{Z}|\times n}$ and define the two character matrices
\begin{align}\label{eq:zd_F_matrices}
    F_p &:= \exp\!\Big(\tfrac{2\pi i}{d}\,K\,X^T\Big)\in\mathbb{C}^{|\mathcal{K}|\times|\mathcal{X}|}, \\ \qquad F_q &:= e^{iE}\in\mathbb{C}^{|\mathcal{K}|\times|\mathcal{Z}|},
\end{align}
where $E$ is the accumulated-phase-difference matrix of Sec.~\ref{sec:efficient_software} built from $(K,Z)$ with $M = \boldsymbol{0}$. Writing $\mathbf{s}_p := F_p\,\boldsymbol{1}_{|\mathcal{X}|}$ and $\mathbf{s}_q := F_q\,\boldsymbol{1}_{|\mathcal{Z}|}$ for the row-sum vectors, the unbiased estimator \eqref{eq:mmd_unbiased} then takes the compact form
\begin{multline}\label{eq:mmd_unbiased_matrix}
    \widehat{\mathrm{MMD}^2_{\mathrm{u}}}(\mathcal{X},\mathcal{K},\mathcal{Z},\btheta) = \frac{\|\mathbf{s}_p\|_2^{2}}{|\mathcal{K}|\,|\mathcal{X}|(|\mathcal{X}|-1)}
    + \frac{\|\mathbf{s}_q\|_2^{2}}{|\mathcal{K}|\,|\mathcal{Z}|(|\mathcal{Z}|-1)}\\
    -\frac{2\,\mathrm{Re}\langle\mathbf{s}_p,\,\mathbf{s}_q\rangle}{|\mathcal{K}|\,|\mathcal{X}|\,|\mathcal{Z}|}
    -\frac{1}{|\mathcal{X}|-1} - \frac{1}{|\mathcal{Z}|-1}.
\end{multline}

\end{document}